\documentclass[graybox]{svmult}

\usepackage{graphicx}        
\usepackage{multicol}        
\usepackage[bottom]{footmisc}

\usepackage{newtxtext}       %
\usepackage{newtxmath}       

\usepackage{url} 
\usepackage{booktabs}
\usepackage{braket}

\usepackage{arydshln}

\makeindex             

\begin{document}

\title*{Experiments with interacting \\ ultracold polar molecules}
\titlerunning{Experiments with interacting ultracold polar molecules}
\author{
Philip~D.~Gregory
\and
Jonathan~M.~Mortlock
\and
Daniel~K.~Ruttley
\and
Simon~L.~Cornish
}

\institute{
Philip~D.~Gregory
\at
Department of Physics, Durham University, South Road, Durham, DH1 3LE, United Kingdom
\email{p.d.gregory@durham.ac.uk}
\and
Jonathan~M.~Mortlock
\at
Department of Physics, Durham University, South Road, Durham, DH1 3LE, United Kingdom
\email{jonathan.m.mortlock@durham.ac.uk}
\and
Daniel~K.~Ruttley
\at
Department of Physics, Durham University, South Road, Durham, DH1 3LE, United Kingdom
\email{daniel.k.ruttley@durham.ac.uk}
\and
Simon~L.~Cornish
\at
Department of Physics, Durham University, South Road, Durham, DH1 3LE, United Kingdom
\email{s.l.cornish@durham.ac.uk}
}
\maketitle
\abstract
{Ultracold polar molecules have emerged as an exciting platform for experiments in quantum science, with many proposed applications seeking to leverage the existence of long-range dipolar interactions and the rich internal structure of long-lived states. However, the complexity of molecules also presents many experimental challenges. In this chapter, we review the experimental advances over the past two decades that have brought ultracold molecules under full quantum control. We focus on three key topics: control of ultracold molecular collisions, engineering dipolar interactions and nascent quantum simulations in optical lattices.}

\clearpage

\section{Introduction}

Ultracold polar molecules offer many intriguing properties for applications in quantum science and technology~\cite{Carr_2009}. They possess a rich internal structure arising from the interplay of hyperfine, rotation, vibration, and electronic degrees of freedom, which provides a vast Hilbert space in which to encode quantum information or pseudo spins. Their intrinsic molecular-frame electric dipole moments can be leveraged to engineer controllable long-range anisotropic dipole-dipole interactions between molecules and to drive fast microwave transitions between rotational states of opposite parity. Crucially, dipole-dipole interactions are accessible in long-lived rotational states in the electronic ground state, allowing long evolution times in quantum many-body settings. Given this unique combination of properties, it is not surprising that many applications have been proposed for ultracold polar molecules spanning the fields of quantum-state-controlled chemistry~\cite{Bell2008, Quemener2012, Balakrishnan2016, Karman2024}, quantum simulation~\cite{Barnett2006,Micheli2006,Capogrosso-Sansone2010,Pollet2010,Gorshkov2011,Gorshkov2011b,Zhou2011,Hazzard2013,Zoller2013,Sundar2018,Sundar2019,Feng2022,Cohen2022} and computation~\cite{DeMille2002,Yelin2006,Pellegrini2011,Wei2016,Ni2018,Sawant2020,Hughes2020,Zhang2022Rydberg,Wang2022,Asnaashari2023}, and the precision measurement of fundamental constants~\cite{Salumbides2013,Safronova2018,ACME,Hutzler2020,Roussy2023,Arrowsmith-Kron2024,DeMille2024}.

For a long time, it was not obvious how to bring molecules into the ultracold regime. Many early efforts explored slowing cold molecular beams generated by supersonic expansion, for example, using Stark decelerators that harness the interaction between polar molecules and electric fields~\cite{Stuhl2014}. This approach led to novel developments such as the centrifuge decelerator~\cite{Chervenkov2014,Wu2017}, the loading of cold polar molecules into electrostatic traps~\cite{Meerakker2005,Rieger2005,Kleinert2007,Hogan2009,Englert2011} and Sisyphus cooling techniques applicable to a variety of polar molecules~\cite{Zeppenfeld2012}. Ultimately, however, two alternative methods emerged for the production of molecules deep in the ultracold regime: molecules may either be assembled from pre-cooled ultracold atoms, or hot molecules formed by chemical processes may be laser-cooled to ultracold temperatures. These two techniques are applicable to different molecular species and present different advantages and challenges, such that the two approaches are complementary.

The production of ultracold molecules by association exploits the simplicity of atomic laser cooling techniques, and has been successfully applied to various bialkali combinations. However, the species of molecules is limited to those whose constituent atoms are amenable to laser cooling, and the pathway for association can sometimes be complex. The first experiments to prepare ultracold polar molecules in the electronic and vibrational ground state used photoassociation~\cite{Nikolov2000, Sage2005, Deiglmayr2008}, whereby molecules are formed using optical resonances in a magneto-optical trap~\cite{Jones2006}. However, this approach has difficulty in coherently preparing the molecules in a single hyperfine level of the ground state.

The most successful technique for associating atoms is a two-step coherent process that starts with an ultracold mixture of ground-state atoms in an optical trap. The first step is magnetoassociation on an interspecies Feshbach resonance. A Feshbach resonance occurs when a weakly-bound molecular state becomes close in energy to the energy of free atoms, and a resonant coupling between these states gives rise to an avoided crossing~\cite{Chin2010}. By sweeping the magnetic field down across the Feshbach resonance, pairs of atoms can adiabatically follow the avoided crossing to form weakly-bound molecules. Secondly, these molecules are transferred to their absolute ground state using stimulated Raman adiabatic passage (STIRAP)~\cite{Vitanov2017} by driving optical transitions in the molecule. Both of these steps are adiabatic so that the temperature of the resulting molecular gas mirrors that of the original atomic mixture. Moreover, the method prepares the molecules in a single hyperfine level of their electronic, vibrational, and rotational ground state; the ideal starting point for many experiments. This technique has been demonstrated in a range of polar~\cite{Ni2008,Takekoshi2014,Molony2014,Park2015,Guo2016,Seesselberg2018,Yang2019,Hu2019,Voges2020,Cairncross2021,Stevenson2023,He2024} and non-polar~\cite{Danzl2008} bialkali molecules with $^1\Sigma$ ground states, and in the metastable $a^3\Sigma$ state~\cite{Lang2008,Rvachov2017}. In these experiments, detection is typically performed by reversing the association process thereby breaking the molecules apart into their constituent atoms; this process is hyperfine-state sensitive, such that only molecules in the particular state that is populated by the STIRAP are detected.

Direct laser cooling of molecules trades the preparation of ultracold atomic mixtures and the need to find a coherent association pathway for more complex laser-cooling schemes. Laser cooling typically requires a `closed' optical transition where the particle returns back to the same initial state after optical excitation, thereby allowing the particle to repeatedly scatter photons indefinitely. In this context, the extra degrees of freedom associated with vibration and rotation of the molecule make finding closed transitions difficult. Fortunately, rotational branching can be eliminated by driving transitions from $N=1$ in the ground state to $N'=0$ in the excited state \cite{Stuhl2008}; here $N$ is the quantum number for rotational angular momentum and the selection rule $\Delta N = \pm 1$ only permits decay back to $N=1$. However, transitions between vibrational states do not have such selection rules. Instead, transition rates are dictated by the square of a wavefunction overlap integral, known as a Franck-Condon factor, which means that an electronically-excited molecule may decay down to any number of vibrational levels of the ground electronic state and are therefore not straightforwardly closed. 

However, Di Rosa highlighted that it is possible to identify species of molecules that possess favourable Franck-Condon factors such that decay is only likely to one of a small number of vibrational states \cite{DiRosa2004}. By introducing a suitable number of repump lasers, these molecules can scatter enough photons to be confined in a magneto-optical trap before being lost to dark states. Direct laser cooling has the advantage of being applicable to different species of molecules, perhaps most notably SrF~\cite{Shuman2010,Barry2014,Norrgard2016}, CaF~\cite{Truppe2017,Anderegg2017,Anderegg2018,Cheuk2018,Caldwell2019,Lu2022} and YO~\cite{Collopy2018,Wu2021,Burau2023} which each have $^2\Sigma$ ground states and therefore both electric and magnetic dipole moments; where the magnetic moment comes from the presence of an unpaired electron. Direct laser cooling has also been recently applied to BaF \cite{Zeng2024,Rockenhauser2024}, the $^1\Sigma$ AlF molecule~\cite{Padilla-Castillo2025}, and to larger polyatomic molecules such as CaOH~\cite{Vilas2022}. 

Despite the enormous success in producing ultracold molecules, many challenges still had to be overcome to be able to exert sufficient control over the molecules to unlock the exciting applications in quantum science. These challenges include coherent control of the internal state in the presence of applied fields (magnetic, optical and electric), engineering dipolar interactions, producing ordered arrays of molecules and controlling molecular collisions. This chapter will describe how each of these challenges has been overcome over the last decade through the collective effort of theorists and experimentalists around the world. We mostly focus on bialkali molecules, as the samples produced in experiments are generally colder and denser, bringing many applications and problems within closer reach. The structure of the chapter is as follows. We begin with a brief account of the structure and interactions of molecules, and the methods developed for coherent internal state control. We then describe the efforts to understand and control ultracold molecular collisions, culminating in a discussion of collisional shielding. The later sections describe recent advances using optical lattices and optical tweezers to implement controlled dipolar interactions between molecules. Finally, we conclude with a short outlook on the prospects for the field.

\section{The structure and interactions of polar molecules}

\subsection{Rotational and hyperfine structure} \label{sec:structure:structure}

Many of the proposed applications of ultracold molecules utilise the rotational degree of freedom, and so an understanding of the rotational structure is therefore essential. For simple diatomic molecules, the rigid-rotor model~\cite{BrownandCarrington} gives sufficient insight. In this model, the rotational eigenstates $\ket{N,M_N}$ are labelled by the rotational angular momentum $N$ and its projection $M_N$ along the quantisation axis. The energies of the rotational states, given by $E_N = B N(N+1)$ where $B$ is the rotational constant, form a ladder with anharmonic spacing. Electric-dipole transitions between rotational states follow the selection rules $\Delta N=\pm1$ and $\Delta M_N=0,\pm1$ and typically lie in the microwave domain; the anharmonic spacing of states ensures that specific rotational transitions can be addressed with high selectivity, while the low transition frequencies give rise to negligible decay, even from highly-excited rotational states. These states are therefore ideal for encoding spins or storing quantum information.

\begin{figure}[t]
    \centering
    \includegraphics[width=\textwidth]{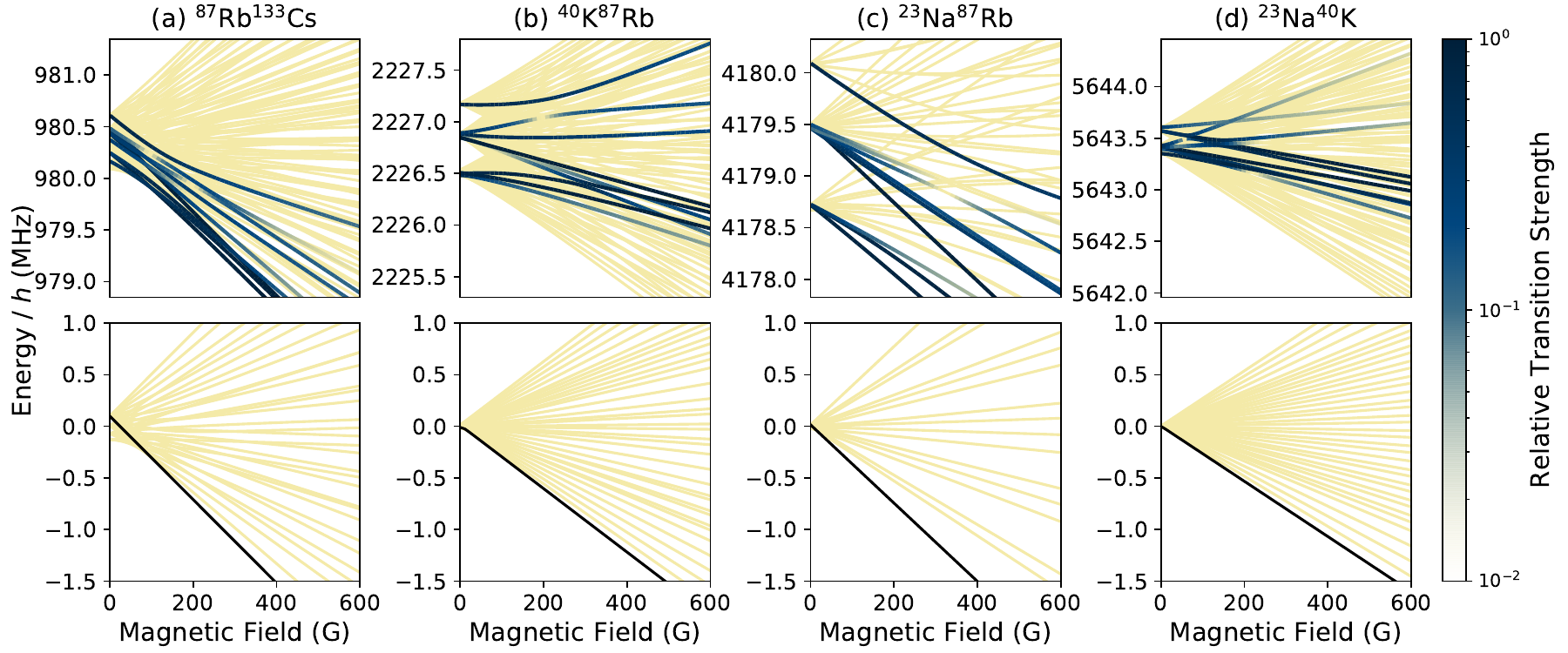}
    \caption{Zeeman structure of states with $N=0, 1$ for~(a)~$^{87}$Rb$^{133}$Cs, (b)~$^{40}$K$^{87}$Rb, (c)~$^{23}$Na$^{87}$Rb, (d)~$^{23}$Na$^{40}$K. In each of the lower panels there is one line coloured black; this is the $N=0$ state that becomes the hyperfine ground state at large magnetic field. In the upper panels, the colouring of the states is codified by the relative transition strength from the hyperfine ground state. \textit{Source: (a, b, c, d) adapted from Ref.~\cite{Blackmore2023}.}}
    \label{fig:Zeeman}
\end{figure}

In this model the rotational levels are $(2N+1)-$fold degenerate. However, in practice, there is hyperfine structure associated with the nuclear spins in the molecule and the degeneracy with respect to $M_N$ is lifted by an applied magnetic field. Example structures for various bialkali molecules are shown in Fig.~\ref{fig:Zeeman}. For bialkali molecules in particular, the hyperfine states often form a dense structure. Mixing between uncoupled basis states can lead to many allowed transitions between any given pair of rotational states. This mixing varies with the strength of the applied magnetic (or electric) field, allowing the relative strengths of the allowed transitions to be smoothly varied~\cite{Hermsmeier2024}. Coherent control of the rotational state of polar molecules is typically performed using microwave pulses to coherently drive transitions between rotational states~\cite{Ospelkaus2010control,Will2016,Gregory2016,Guo2018,Williams2018,Blackmore2020pccp}. This transfer can be made highly efficient due to the lack of spontaneous emission and the availability of low-noise microwave sources.

\subsection{Engineering dipolar interactions} \label{sec:structure:interactions}

Rich physics can emerge in systems of ultracold polar molecules due to long-range and anisotropic dipole-dipole interactions between molecules. Diatomic molecules exhibit an electric dipole moment when the two atoms forming the molecule possess different electronegativities. In such cases, the electric dipole moment then points along the internuclear axis of the molecule. Generally, the dipole-dipole interaction between a pair of molecules is described by the Hamiltonian \cite{Wall2015}
\begin{equation} \label{eq:structure:dipolar-hamiltonian}
    H_\mathrm{dd} = \frac{\vec{d_1}\cdot\vec{d_2}-3(\vec{d_1}\cdot\vec{\hat{r}})(\vec{d_2}\cdot\vec{\hat{r}})}{r^3},
\end{equation}
where $r$ and $\vec{\hat{r}}$ are the magnitude and direction of the intermolecular vector and $\vec{d_i}$ is the dipole operator of molecule $i$.
The direction in which the dipole is pointed depends on the orientation of the molecule, and this is inextricably linked to its rotational state. 
However, for molecules prepared in any single rotational state, the wavefunction that describes the orientation has symmetry such that the molecule has no preferred orientation and the dipole moment in the laboratory frame is zero. 

Dipolar interactions must therefore be carefully engineered in experiments. Several different approaches can be used. The application of external fields can break the wavefunction symmetry, thus inducing an electric dipole moment in the laboratory frame. Perhaps the simplest approach is to apply a DC electric field to the molecule (see chapter (10.2)). This mixes rotational state components with the same $M_N$, such that the wavefunction becomes asymmetric and the dipole points preferentially either parallel or anti-parallel to the field. Similarly, AC electric fields in the microwave domain can be used to dress the rotational states (see chapter (10.2)). Alternatively, molecules may be prepared in quantum superpositions of states, resulting in dipoles that oscillate or rotate in time, again leading to dipolar interactions. 
Finally, a pair of molecules prepared in neighbouring rotational states that are connected by an allowed electric-dipole transition can resonantly exchange the excitation, leading to so-called spin-exchange (or flip-flop) interactions. Here, the transition dipole moment dictating the strength of the interaction can be controlled through the choice of hyperfine component and the applied magnetic field~\cite{Hermsmeier2024}.
As we will see, all of these approaches to engineering dipolar interactions have their place in current experiments.

\subsection{Trapping molecules} \label{sec:structure:trapping}

The majority of experiments that have been performed with ultracold molecules trap them in optical potentials. For bialkali molecules in the electronic ground state (the $\mathrm{X}^1\Sigma^+$ potential), this is necessary as the molecules do not possess an electronic magnetic moment and therefore cannot be easily trapped magnetically. Furthermore, many applications exploiting dipolar interactions require $\upmu$m\,-scale separations, comparable to common trapping wavelengths, making optical potentials the natural choice for manipulation of ultracold molecules. The optical trapping potential depends on the polarisability of the molecule $\alpha$ and the laser intensity $I$ as $\alpha I/ 2\epsilon_0 c$, where $\epsilon_0$ is the permittivity of free space and $c$ is the speed of light. The polarisability exhibits many poles due to the abundance of transitions~\cite{Vexiau2017} and trapping wavelengths must be chosen with care.

Critically, due to their anisotropic structure, different rotational states of a molecule generally have markedly different polarisabilities. This means that rotational transitions for trapped molecules
experience large differential light shifts. This causes rapid dephasing of rotational state superpositions and suppresses \emph{resonant} spin-exchange interactions. To understand this, it is instructive to consider the polarisability for a diatomic molecule in more detail. 
In the molecule frame, the polarisability can be decomposed into components parallel ($\alpha_\parallel$) and perpendicular ($\alpha_\perp$) to the internuclear axis. These polarisabilities derive from molecular transitions with different symmetries~\cite{Vexiau2017}.
Then, the polarisability $\alpha$ of a molecule with internuclear axis at angle $\theta$ to the quantisation axis becomes $\alpha(\theta)=\alpha_\parallel\cos^2\theta+\alpha_\perp\sin^2\theta \equiv \alpha^{(0)} + \alpha^{(2)}(3\cos^2\theta-1)/2$. Here, $\alpha^{(0)}\equiv(\alpha_\parallel+2\alpha_\perp)/3$ is the isotropic polarisability and $\alpha^{(2)}\equiv2(\alpha_\parallel-\alpha_\perp)/3$ is the anisotropic polarisability. 
For the spherically symmetrical ground state $\ket{0,0}$, the anisotropic polarisability has no effect.
However, for rotationally excited states, the polarisability operator mixes hyperfine states within a rotational manifold. The light shift of a particular state is therefore found by evaluating the matrix elements $\braket{N',M_N'|\alpha|N,M_N}$.
Explicitly, for the first-excited manifold,
\begin{equation}
    \braket{N',M_N'|\alpha|N,M_N} \propto \alpha^{(0)} + \frac{\alpha^{(2)}}{5}\begin{pmatrix}
        -P_2(\cos\theta) & +\frac{3}{\sqrt2}\sin\theta\cos\theta & +\frac{3}{2}\sin^2\theta \\
        +\frac{3}{\sqrt2}\sin\theta\cos\theta & 2P_2(\cos\theta) & -\frac{3}{\sqrt2}\sin\theta\cos\theta \\
        +\frac{3}{2}\sin^2\theta & -\frac{3}{\sqrt2}\sin\theta\cos\theta & -P_2(\cos\theta)
    \end{pmatrix},\label{eq:lightshift}
\end{equation}
in the basis $\{\ket{1,-1},\ket{1,0},\ket{1,+1}\}$ and where $P_2(\cos\theta) = (3\cos^2\theta-1)/2$.

Overcoming the strong differential light shifts that result from these tensor light shifts has been a major challenge for the field. In experiments, molecules often experience different intensities (for example, when confined in an optical trap). Then the differential light shifts lead to dephasing and an observed loss of coherence for quantum-state superpositions. However, the form of Eq.~\eqref{eq:lightshift} elucidates two approaches that have been successful in minimising dephasing, and thereby maximising the coherence time of rotational state superpositions. The first uses light at a ``magic angle'' to eliminate first-order shifts by setting \mbox{$P_2(\cos\theta)=0$}~\cite{Kotochigova2010}.
The second uses light at a ``magic wavelength'' where $\alpha^{(2)}$ is eliminated to remove light shifts to first and second order~\cite{Guan2021}. Details of the experimental implementation of these approaches are presented in Sec.~\ref{sec:coherences}.

\subsection{A comparison of molecules}

We end this section by comparing in Table~\ref{Tab:1} the key properties of laser-cooled diatomic molecules and bialkali molecules that are of current experimental interest. The values in this table serve as a reference for later sections of the chapter and highlight the broad range of interaction strengths achievable in experiments. We note that a new range of experiments seeks to create silver-alkali molecules where the dipole moments and hence interactions are even larger~\cite{TomzaSilver,Vayninger2025}.

\begingroup
\setlength{\tabcolsep}{10pt} 
\renewcommand{\arraystretch}{1.5}
\begin{table}
    \centering
    \begin{tabular}{cccccc}
        \toprule
       Species  & Electronic & EDM, & $2B$ & Critical $E$ field & Interaction ($V_{\mathrm{dd}}$) \\
         & state  &  $d$ (D) & (GHz) & (kV/cm) &  at 1$\upmu$m (Hz) \\
       
        \midrule
        
        LiK &$\mathrm{X}^1\Sigma$ & 3.45 & 15.6 & 4.5 & 2400 \\
        
        LiRb &$\mathrm{X}^1\Sigma$ & 4.00 & 13.2 & 3.3 & 3220 \\
        
        LiCs &$\mathrm{X}^1\Sigma$ & 5.52 & 11.6 & 2.1 & 6130 \\
        
        NaK & $\mathrm{X}^1\Sigma$& 2.72 & 5.64 & 2.1 & 1490 \\
        
        NaRb & $\mathrm{X}^1\Sigma$& 3.10 & 4.18 & 1.3 & 1930 \\
        
        NaCs & $\mathrm{X}^1\Sigma$& 4.75 & 3.56 & 0.74 & 4540 \\
        
        KRb & $\mathrm{X}^1\Sigma$& 0.57 & 2.23 & 3.9 & 66 \\
        
        KCs & $\mathrm{X}^1\Sigma$& 1.91 & 1.86 & 0.97 & 730 \\
        
        RbCs & $\mathrm{X}^1\Sigma$ & 1.23 & 0.98 & 0.81 & 300 \\
        NaLi & $\mathrm{a}^3\Sigma$ & 0.18 & 8.2 & 45 & 6.5 \\
        \hdashline
        SrF & $\mathrm{X}^2\Sigma$ & 3.47 & 7.5 & 2.1 & 2400\\
        CaF & $\mathrm{X}^2\Sigma$ & 3.07 & 10.27 & 3.3 & 1900 \\
        BaF &  $\mathrm{X}^2\Sigma$ & 3.17 & 6.48 & 2.0 & 2020 \\
        YO & $\mathrm{X}^2\Sigma$ & 4.50 & 23.263 & 5.1 & 4070\\
        AlF & $\mathrm{X}^1\Sigma$ & 1.50 & 33.0 & 22 & 450\\
        
        \bottomrule
    \end{tabular}
    \caption{Comparison of the key properties of molecules of current experimental interest. Entries above the dashed line correspond to molecules produced by association of atoms, while those below the line may be directly laser cooled. All properties are given for molecules in the vibrational ground state of the electronic state that is listed. The quantity $2B$ is twice the rotational constant, corresponding to the transition frequency from $N=0 \rightarrow 1$. The critical electric ($E$) field is equal to $B/d$, where $d$ is the molecular-frame electric dipole moment given in the third column, and when applied to the molecule generates a lab-frame dipole of $d/3$. The interaction assumes a transition dipole moment of $d/\sqrt{3}$ and the maximum angular factor. Note that the properties of NaLi are given for the triplet~$^3\Sigma$ rather than the singlet~$^1\Sigma$ ground state due to relevance to experiments in Ref.~\cite{Rvachov2017}.}
    \label{Tab:1}
\end{table}
\endgroup

\section{Understanding and controlling ultracold molecular collisions}\label{sec:3}

Understanding and controlling ultracold \emph{atomic} collisions has been critical to the development of the field of quantum gases and has enabled the study of a vast range of physical phenomena. The use of Feshbach resonances to control atomic interactions has been fundamental~\cite{Chin2010}. Extending the same level of control to ultracold molecules will be essential to many proposed applications. However, as experiments have confirmed, collisions between ultracold molecules are far more complex and far harder to understand theoretically. In this section, we review the early observations of ultracold molecular collisions in thermal gases and show how these all pointed to strong loss during the collision. Then, we describe how dipole-dipole interactions between molecules can be engineered to prevent such lossy collisions, leading to the control of molecular collisions and the creation of quantum-degenerate samples of polar molecules through evaporative cooling.

\subsection*{Exothermic reactions and sticky collisions}

Early bulk gas experiments with ultracold polar molecules were performed in pioneering experiments in the group of Jin and Ye using fermionic KRb. They prepared a thermal sample of molecules in an optical dipole trap and observed collisional loss of molecules from the trap as a function of time~\cite{Ospelkaus2010collisions}. When the sample was prepared in a single quantum state, the lowest-order collisions that are allowed are $p$-wave, and as such the loss scales linearly with temperature. They were also able to study the rate of temperature-independent $s$-wave collisions by preparing the molecules in a 50:50 mixture of different hyperfine states. The $s$-wave and $p$-wave loss rates they observed were both fast and comparable to their respective two-body universal limit; this is the limit where there is unity probability of loss whenever molecule pairs approach short range~\cite{Idziaszek2010}.

\begin{figure}[t!]
\centering
\includegraphics[width=1.\columnwidth]{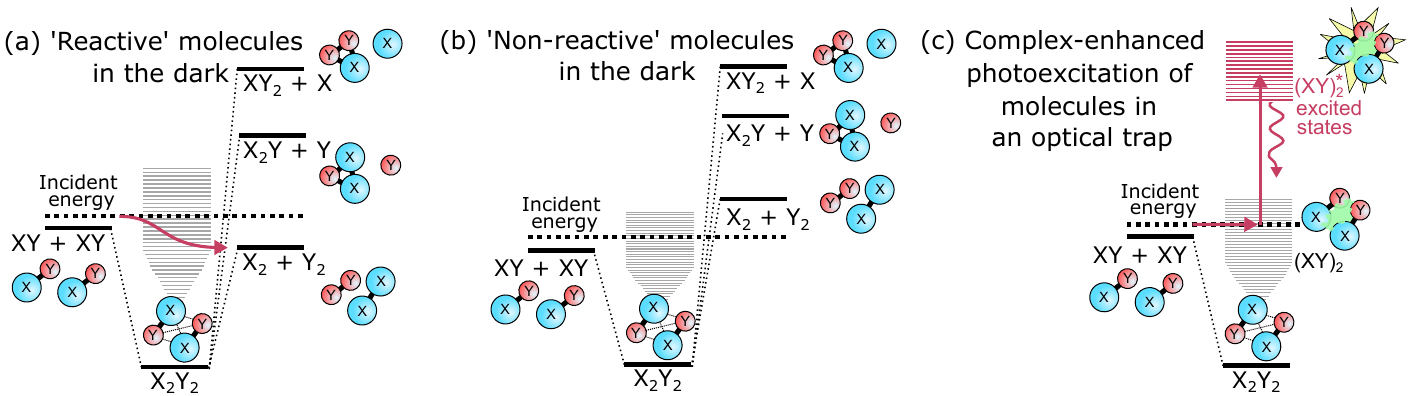}
\caption{The relevant energy levels for collision dynamics involving pairs of (a)~reactive and (b)~non-reactive diatomic molecules~(XY). In each case, the molecules collide with some incident energy and must navigate through the molecular potential and near-resonance states associated with the 4-atom X$_2$Y$_2$ molecule before exiting the collision. In (c), we show the proposed mechanism for loss of molecules in optical traps whereby the molecules become stuck together in a temporary collision complex (XY)$_2$ which then experiences rapid photoexcitation to, and subsequent spontaneous emission from, the electronically excited states of the 4-atom molecule.  }
\label{fig:ReactionPathways}
\end{figure}

Fast collisional loss presents a problem for many experiments. Evaporative cooling techniques for example rely on a large fraction of the collisions being elastic in order for the sample to remain in thermal equilibrium. The source of fast loss for KRb was initially determined to be due to pairs of molecules undergoing the exothermic atom-exchange reaction
$\mathrm{KRb}+\mathrm{KRb}\rightarrow \mathrm{K}_2 + \mathrm{Rb}_2$ following the path illustrated in Fig.~\ref{fig:ReactionPathways}(a). While potentially unavoidable for KRb, it was shown theoretically that such reactions, along with all other atom-exchange reactions, were energetically forbidden for other bialkali combinations, with the exception of those containing Li~\cite{Zuchowski2010}. For these species, the relevant energy level structure is shown in Fig.~\ref{fig:ReactionPathways}(b). In this case the reactive channels are inaccessible and although the colliding molecules explore a high density of resonant states associated with the (XY)$_2$ complex, the expectation is that they should return to the separated molecules following the collision. However, later collisional studies with nonreactive RbCs~\cite{Takekoshi2014,Gregory2019}, NaRb~\cite{Ye2018}, NaK~\cite{Park2015, Bause2021}, and NaCs~\cite{WarnerThesis} have all found fast losses close to the universal limit. Near-universal loss rates have also been observed in triplet ground-state NaLi~\cite{Rvachov2017}, and for non-alkali molecules; such as between pairs of CaF molecules that were prepared in optical tweezers following direct laser cooling~\cite{Cheuk2020}. The study in NaRb~\cite{Ye2018} compared the loss rates for ground-state and vibrationally-excited molecules, thereby controlling the reactivity of the molecules, and found that the observed loss rate was independent of the reactivity. A different mechanism was therefore required to explain the collisional loss of these new `nonreactive' molecules; with a two-step process proposed and illustrated in Fig.~\ref{fig:ReactionPathways}(c). 

In the first step, two of these molecules collide with each other, but rather than the collision being instantaneous, the collision instead takes some time due to the presence of many near-resonant molecular states~\cite{Mayle2012,Mayle2013}. The two molecules in effect become stuck together for a time while they explore the potential energy surface associated with the larger four-atom molecule. This first step is commonly referred to as a `sticky' collision, and the temporary structure of the two molecules stuck together is a collision complex. In the limit where the spacing of the resonances is much larger than the width, then the motion during the collision is ergodic and the lifetime of the complex can be estimated using Rice-Ramsperger-Kassel-Marcus (RRKM) theory~\cite{Levine2005}. The formation of the collision complex alone does not cause permanent loss as the complex will eventually decay back to free molecules. 

The second step causes removal of the complex from the trap, or alternatively, prevents the complex from naturally decaying back to free molecules. Calculations by Christianen~\emph{et al.} for the $(\mathrm{NaK})_2$ complex suggested that light from the optical trap may be resonant with electronic transitions available to the complex over a broad wavelength range~\cite{Christianen2019}. This would lead to the complex quickly scattering photons, and likely cause the molecule pair to become lost before it can decay. 

\begin{figure}[t!]
\centering
\includegraphics[width=1.0\columnwidth]{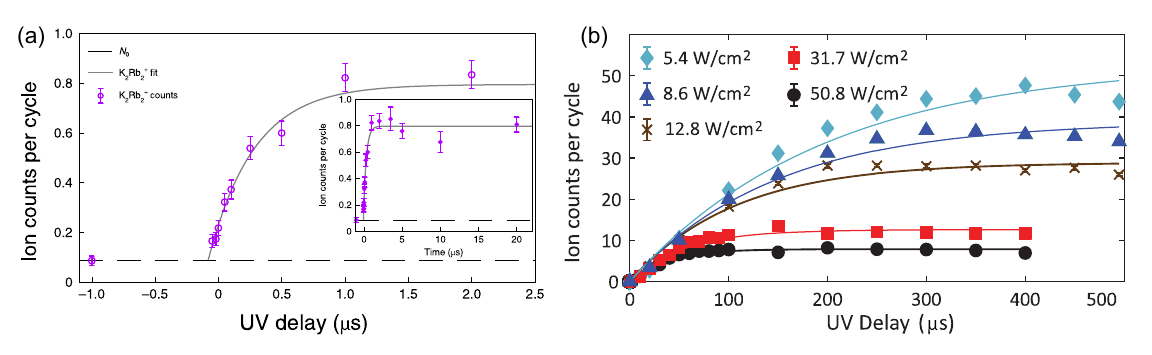}
\caption{Direct measurements of the lifetime of collision complexes for (a)~KRb+KRb and (b)~KRb+Rb. In each case, the $x$~axis indicates a time delay between the off-resonant trap light being either (a)~switched off completely or (b)~suddenly reduced to a lower level and the detection of the complexes via an ionising ultraviolet pulse. For (a), the inset shows the same results but over a larger range of UV delay times. For (b), the level to which the trap light is reduced is indicated by the symbols. \textit{Source: (a) is adapted from Ref.~\cite{Liu2020}, (b) is adapted from Ref.~\cite{Nichols2022}.} }
\label{fig:Ni_Intermediates}
\end{figure}

The only experiments to directly detect the presence of ultracold collision complexes have been performed with reactive KRb. In pioneering experiments in the Ni group, an ultracold gas of KRb molecules is prepared and then the components of the gas photoionised and detected using a combination of mass spectrometry and velocity-map imaging. This allows the precise detection of all species present in the sample, not just molecules prepared in one specific quantum state. Using this approach they have directly observed the formation of the $(\mathrm{KRb})_2$ complex~\cite{Hu2019} which is formed as a part of the energetically allowed atom-exchange reaction. Perhaps surprisingly, they also found that photoexcitation of the complex was the dominant process undergone in an optical trap even for this reactive molecule. This indicates that the rate at which the complexes are excited is faster even than the energetically-allowed chemical reaction. By turning the trap light off, they were able to detect the time-dependent variation in the number of collision complexes present as a function of the dark time as shown in Fig.~\ref{fig:Ni_Intermediates}(a)~\cite{Liu2020}. From this, they measure the lifetime of the collision complex in the dark to be $\tau_\mathrm{c}^\mathrm{KRb}=0.36(3)$\,$\upmu$s, within a factor of 2 of the RRKM prediction~\cite{Christianen2019dos}. With the trap light on, they found that the complexes typically live for on the order of 10\,ps or less before they are photoexcited.

This novel detection method that enables measurement of reactants, transients, and products of reactions has allowed new insights into how quantum coherence may be conserved through these atom-exchange reactions. The Ni group have shown that they are able to control the quantum state occupied by the $\mathrm{K}_2$, $\mathrm{Rb}_2$ products by varying the state of the initial $\mathrm{KRb}$ reactants~\cite{Liu2021,Hu2021} and, by preparing the reactants in an entangled state, that nuclear spin entanglement can be conserved through the reaction~\cite{Liu2024}.  

Studies of complex formation in nonreactive molecules have been more difficult as the existing experiments are only able to detect the initial molecule species in one specific quantum state. However, an indirect approach to detecting these complexes has been implemented. This detection scheme requires modulation of the intensity of the optical trap, such that the molecules experience a time-averaged potential and are periodically placed in the dark. During the dark time, molecules that form collision complexes are no longer lost, and as such there should be a small suppression of the loss during each dark time that can build up to a large change in the number of molecules over many modulation cycles when compared with a trap that does not go dark. By varying the frequency of the modulation, we can vary the duration of each dark time; thereby enabling time-resolved detection that allows indirect measurement of the lifetime of the collision complexes. 

This scheme was first implemented for RbCs molecules and yielded a measurable suppression of loss associated with the molecules being held in the dark. From these experiments, the lifetime of the complex in the dark was measured to be $\tau_\mathrm{c}^\mathrm{RbCs}=0.53(6)$\,ms~\cite{Gregory2020}. As with the KRb result, this is within a factor of 2 of the RRKM prediction~\cite{Christianen2019}. However, so far this approach has failed to detect measurable suppression for any other nonreactive molecule, despite these molecules being expected to have much shorter complex lifetimes than the heavy RbCs molecule; shorter lifetimes should be easier to detect using this approach as the longest available dark time is limited by the need to modulate the intensity much faster than the time-averaged trap frequency. Studies of bosonic molecule collisions using this approach have been published for NaK and NaRb~\cite{Gersema2021}, and results for NaCs presented in~\cite{WarnerThesis}. A study of fermionic NaK was performed by Bause~\emph{et~al.} using this intensity-modulation approach and also by placing the molecules in a blue-detuned box trap~\cite{Bause2021}. The measurements performed with these other nonreactive molecules suggest complex lifetimes many orders of magnitude larger than predicted by RRKM. There is therefore still some mystery surrounding the mechanisms for collisional losses of ultracold molecules that remains to be understood. 

A recent exciting development is the observation of a magnetically-tuned Feshbach resonance in collisions of  NaLi molecules in the rovibrational ground state of the a$^3\Sigma$ potential~\cite{Park2023MoleculeMoleculeResonance}. While near-universal loss is observed at most magnetic fields, Park~\textit{et al.} observed a $\sim25$\,mG wide Feshbach resonance with an associated increase in the loss rate by more than 2 orders of magnitude at its peak. The magnetic field at which the resonance occurs corresponds to where the energy of the colliding molecules becomes degenerate with another open channel, suggesting that the enhancement of loss is associated with a resonant transfer of population between free molecular states. They analyse their results with a model that includes complex formation and extract an estimate of the complex lifetime of 320\,ns, though the lifetime could be much longer due to thermal effects. This work suggests a new way to probe ultracold molecular systems and opens up the prospect of coherent control of the internal states of molecules through collisions.

\subsection*{Atom-molecule collisions and trimer formation}

The study of collisions between atoms and molecules is another exciting frontier for ultracold chemistry. Moreover, these types of collisions are experimentally straightforward to access in experiments where molecules are produced by association; it simply requires only removing one of the constituent atomic species after the association, rather than both. As with molecule-molecule collisions, there are reactive and nonreactive combinations, this time relating to chemical reactions of the type $\mathrm{XY}+\mathrm{X}\rightarrow\mathrm{X}_2+\mathrm{Y}$, and collisions may still be `sticky' though in general with a much shorter sticking time expected due to the much smaller density of states.

Collisions between atoms and molecules with energetically-allowed exchange reactions have generally observed fast collision loss at close to the universal limit~\cite{Ospelkaus2010collisions,Gregory2021collisions, Voges2022}. For nonreactive collisions, however, variation has been observed. The loss rates associated with collisions involving $\mathrm{KRb}+\mathrm{Rb}$ have been measured to be consistent with the universal limit and the same loss mechanism of photoexcitation of collision complexes identified as the leading cause of loss. Nichols~\emph{et~al.} measured the lifetime of the collision complex using the photoionization detection method, as shown in Fig.~\ref{fig:Ni_Intermediates}(b), and found the lifetime of the complex to be 0.39(6)\,ms; a surprising result as it is 5 orders of magnitude larger than the RRKM prediction. Near-universal loss has also been observed for nonreactive $\mathrm{RbCs}+\mathrm{Cs}$, though experiments in an intensity-modulated potential were unable to detect the formation of collision complexes~\cite{Gregory2021collisions}. 

Recent observations by Liu~\emph{et~al.} indicate the possible transfer of energy between the hyperfine energy of a Rb atom to the rotational energy of a KRb molecule during a collision~\cite{Liu2025}. Prior to collision, the Rb atoms were prepared in their upper hyperfine state $(f=2, m_f=2)$, and the KRb molecules in their rotational ground state. After the collision the atom was transferred to the (1,1) ground state, while the molecules populated various states up to and including the $N=2$ rotationally excited state. The possibility of energy transfer between hyperfine and rotational degrees of freedom is surprising and may be connected to the observed long-lifetimes of some collision complexes, as couplings between nuclear spin and rotational angular momentum may drastically increase the number of states that are accessible during a collision. 

Loss rates associated with atom-molecule collisions that were significantly below the universal limit were first observed in collisions between fermionic NaLi, prepared in the triplet ground state, and Na~\cite{Son2020}. When both the molecules and the atoms were prepared in their upper-stretched internal states, the loss rate measured was 2 orders of magnitude below the universal limit. The authors also estimated the ratio of elastic to inelastic collisions to be $\sim300$. By evaporatively cooling the atoms, they showed that this enabled thermalisation between the atoms and molecules leading to highly-efficient sympathetic cooling. 

The loss rates observed in the triplet $\mathrm{NaLi}+\mathrm{Na}$ collisions were highly dependent on the internal states of the atoms and molecules; when the atoms were prepared in their lowest hyperfine state, the collisional loss rate increased such that it was consistent with the universal limit~\cite{Son2020,Rvachov2017}. A similar dependence on the hyperfine state of the atom has been observed in bosonic $\mathrm{NaK}+\mathrm{K}$ collisions. Voges~\emph{et~al.} found the loss rates for this mixture varied by orders of magnitude~\cite{Voges2022}. The lowest loss rate they measured was at least 4 orders of magnitude below the universal limit, and was observed only for $^{39}$K prepared in $(f=1, m_f=-1)$.

Collisions between laser-cooled CaF and Rb atoms have been studied in magneto-optical traps (MOTs)~\cite{Jurgilas2021MOT} and in a magnetic trap~\cite{Jurgilas2021}. In the magneto-optical trap, the atom-molecule collisions limit the lifetime of the molecules with an associated collisional loss rate that is similar to the universal limit. In contrast, the collisional loss is highly dependent on the rotational state in the magnetic trap. For rotationally excited molecules, the loss rate is near universal. However, for molecules prepared in the rotational ground state, the loss rate associated with atom-molecule collisions is below the detection limit for the experiment, independent of their hyperfine state. Their findings suggest that atom-molecule collisions can result in fast rotational relaxation for this mixture. Moreover, as the mixture is in a magnetic trap, collision complexes cannot be photoexcited and, therefore, these studies provide a clean setting in which to explore collisions involving ultracold molecules.

\begin{figure}[t!]
\centering
\includegraphics[width=1.0\columnwidth]{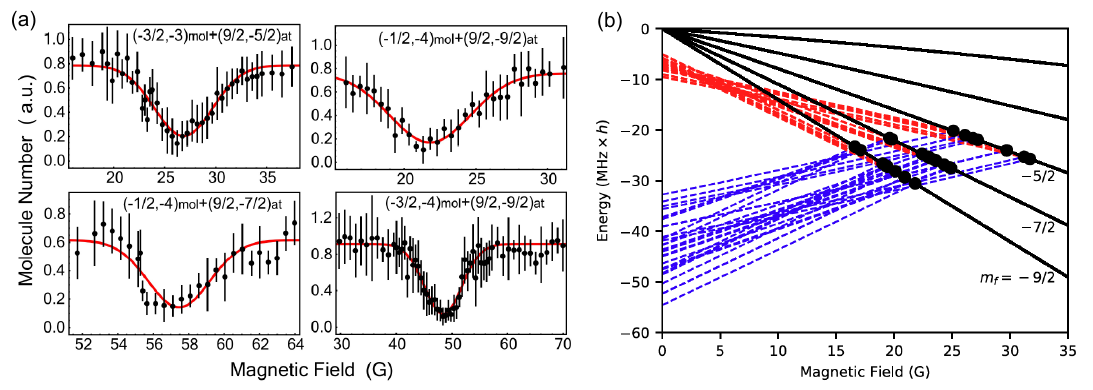}
\caption{Feshbach resonances observed in atom-molecule collisions between NaK and K. (a)~Increased loss of molecules is observed in the experiments as a function of the magnetic field indicating the location of an interspecies Feshbach resonance. Each panel shows results across a different range of magnetic fields. (b)~Two possible explanations (red and blue) for the positions of Feshbach resonances observed below 35\,G. The black solid lines indicate the energy of the free atom-molecule pair, while the red and blue lines show the energies from two different sets of possible NaK$_2$ bound states. Where the bound states are equal to the free pair, a Feshbach resonance may occur. \textit{Source: (a, b) are adapted from Ref.~\cite{Wang2021}.} }
\label{fig:Pan_FeshbachResonances}
\end{figure}

Control of collisions in the ultracold regime is often employed in ultracold \emph{atomic} experiments through the use of magnetically-tunable Feshbach resonances~\cite{Chin2010}. Similar control over atom-molecule collisions has important applications for improved sympathetic cooling and for the formation of triatomic molecules. Feshbach resonances between ultracold atoms and molecules were first observed by Yang~\emph{et~al.} in a mixture of fermionic $\mathrm{NaK}+\mathrm{K}$~\cite{Yang2019}. The Feshbach resonances were identified by a resonant increase in the collisional loss rate as a function of magnetic field as shown in Fig.~\ref{fig:Pan_FeshbachResonances}(a). By comparing the locations of many resonances for atoms and molecules prepared in a range of states, they were able to identify possible quantum numbers of the coupled trimer bound states~\cite{Wang2021}. Two of the possible explanations for resonances observed at low field are shown in Fig.~\ref{fig:Pan_FeshbachResonances}(b). Feshbach resonances have also been observed in collisions between triplet $\mathrm{NaLi}+\mathrm{Na}$~\cite{Park2023}, and the resulting spectrum explained by a theoretical model~\cite{Karman2023}. 

By tuning the magnetic field in close proximity to one of these Feshbach resonances, the ratio of elastic and inelastic two-body collisions can be directly controlled~\cite{Su2022}. This has allowed experiments to use a sympathetic cooling approach to prepare a quantum degenerate mixture of $\mathrm{NaK}$ and $\mathrm{K}$, with elastic collisions enabling thermalisation between the atomic and molecular gases~\cite{Cao2023}. 

The identification of discrete Feshbach resonances in the mixture also opens the door to efficient formation of ultracold samples of triatomic molecules. In a similar approach to how ultracold NaK molecules are formed by association from the ultracold mixture of Na and K atoms using an interspecies Feshbach resonance, the $\mathrm{NaK}+\mathrm{K}$ may be associated to form $\mathrm{NaK_2}$ molecules using these atom-molecule Feshbach resonances. Experiments have successfully produced these triatomic molecules using radio-frequency association~\cite{Yang2022nature} and adiabatic magnetoassociation (i.e. by sweeping the magnetic field across the resonance)~\cite{Yang2022science}. The latter produced a sample of around 4000 weakly-bound $\mathrm{NaK_2}$ molecules with a similar temperature to that of the original mixture. Most recently, Cao~\emph{et~al.} have studied photoassociation in a mixture of fermionic $\mathrm{NaK}+\mathrm{K}$~\cite{Cao2024}. This allows understanding of the excited state structure of the trimer, paving the way for the possible preparation of ground-state $\mathrm{NaK}_2$ either by transfer of the weakly-bound molecules using stimulated Raman adiabatic passage, or via two-photon photoassociation.

\subsection*{Enhancing loss rates with dipolar interactions}

The intermolecular potential that contributes to collisions can be strongly modified by long-range anisotropic dipole-dipole interactions.
Collisions of polar molecules in electric fields were first investigated for KRb in an optical dipole trap by Ni~\emph{et al.}~\cite{Ni2010}. They found a significant increase in the two-body loss rate associated with the introduction of the dipole-dipole interactions and saw evidence that the sample experienced different thermalisation behaviours in parallel and perpendicular to the applied electric field due to the anisotropy of the interaction. A similar increase in the loss rate was also observed for nonreactive NaRb by Guo \emph{et al.}~\cite{Guo2018}. Here they observed a stepwise enhancement of the losses due to the coupling between different partial waves that is induced by the dipole-dipole interactions. 

An alternative approach to introduce dipole-dipole interactions is to prepare the molecules in either a superposition or mixture of different rotational states. For molecules prepared in coherent superpositions of states the molecules can be thought of as oscillating dipoles, where the magnitude of the dipole is proportional to the transition dipole moment that connects the two superposed states. For molecules prepared in either superpositions or mixtures, resonant spin-exchange interactions allow the exchange of rotational angular momentum between the molecules. Collisions in a decohered mixture or rotational states was reported in RbCs~\cite{Gregory2019} with an order of magnitude higher loss rate observed for the mixture compared to molecules prepared in a single rotational state. Resonant dipolar interactions may also be observed by dressing the molecules with a continuously applied microwave field. Yan~\emph{et~al.} studied the effects of preparing Fermionic NaK in microwave dressed states and observed a significant enhancement of the loss rate above the universal limit associated with attractive dipole-dipole interactions~\cite{Yan2020}.

\subsection*{Collisional shielding}

So far we have discussed work in which the introduction of dipole-dipole interactions led to an increase in the rate of collisional loss. However, one of the great advances recently made in this field is to harness these interactions to instead suppress loss. This is achieved by exploiting the repulsive side of the dipole-dipole interaction to engineer a repulsive shield. This shielding prevents the molecules from getting too close to each other during a collision, and therefore stops the formation of complexes that lead to loss. Note, that more details of the theoretical foundation of these shielding mechanisms may be found in the theory chapter (10.2).

The first reports of the suppression of collisional loss with a dc electric field were by de Miranda~\emph{et al.}~\cite{deMiranda2011}. This work, published the year after the study of KRb collisions in~\cite{Ni2010}, placed the molecules in a 1D optical lattice. An electric field was applied that oriented the molecular dipoles along the direction of tight confinement, such that the molecules would approach each other side-by-side. The dipole-dipole interaction in this configuration is repulsive. They observed a suppression of the collisional loss rate by two orders of magnitude when compared with no electric field applied, indicating that there was indeed a mechanism for controlling the rate of collisional loss using these long-range interactions. Valtolina~\emph{et~al.}~\cite{Valtolina2020} applied this technique to colder and denser samples of molecules in the 1D lattice. They measured the ratio of elastic-to-inelastic collisions as $\sim200$. This ratio was sufficient to perform evaporative cooling of the molecules. 

A recent advancement in shielding with dc electric fields is in the use of a resonant effect first observed by Matsuda~\emph{et~al.}~\cite{Matsuda2020}. Here, the molecules are prepared in the rotationally excited state $(N,M_N)=(1,0)$. They observed a strong dependence of the loss rate at a particular electric field, that corresponds to the field needed to make the energy of the two colliding molecules in $(1,0)$ close to degenerate with the energy of a pair of molecules in $(0,0)+(2,0)$. This is due to a dipolar coupling between the two pairs of molecule states that accordingly depends on the separation between the molecules. 

Perhaps surprisingly, this resonant dc electric field shielding also works in 3D~\cite{Matsuda2020, Li2021}. For a carefully chosen electric field there exists a potential energy surface that is repulsive for molecules that first enter the collision either end-to-end or side-by-side. This is possible because the molecular interactions become strong enough to reorient the molecules into their repulsive configuration at the distance over which the shield is effective~\cite{Lassabliere2022}. Experiments performed with KRb molecules demonstrated that this resonant dc electric field shielding suppressed the collisional loss by a factor of 30 and also enabled efficient evaporative cooling to be performed with samples confined in an optical dipole trap~\cite{Li2021}. 

\begin{figure}[t!]
\centering
\includegraphics[width=1.0\columnwidth]{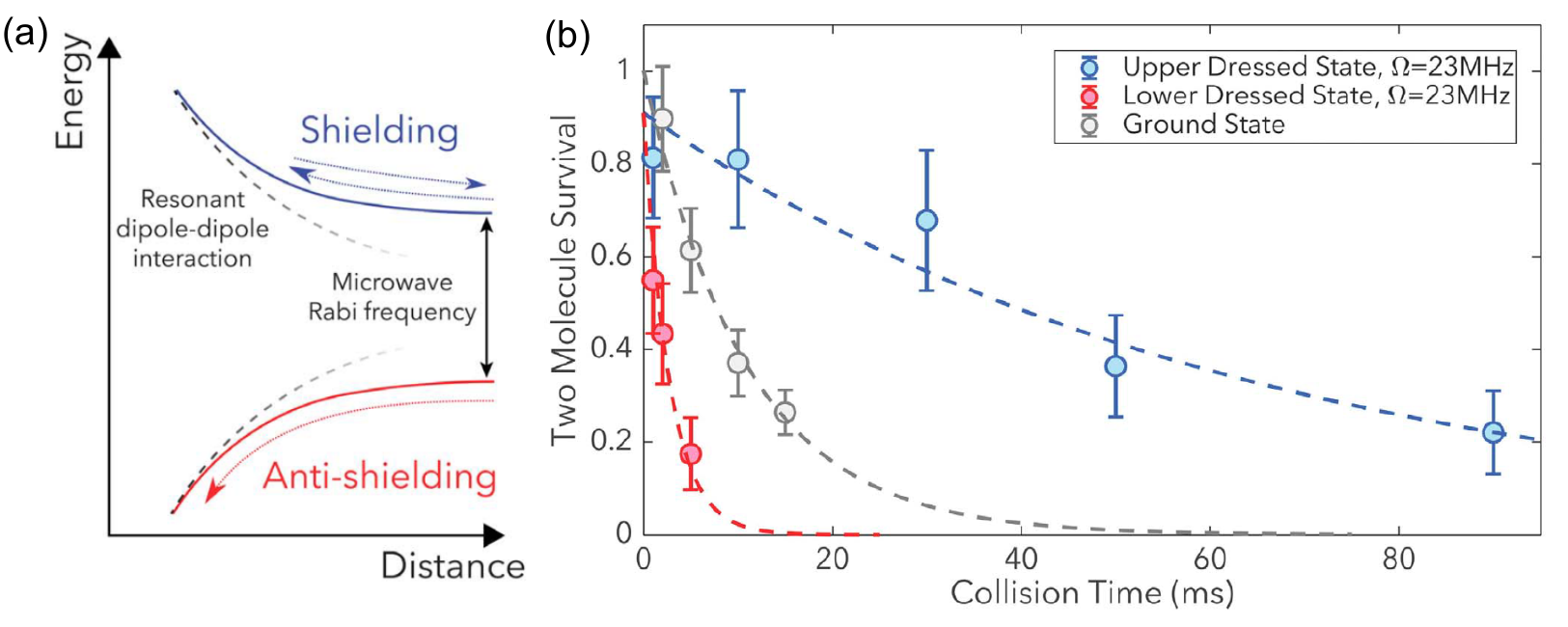}
\caption{Collisional shielding using circularly polarised microwaves, demonstrated between a pair of CaF molecules in the same optical tweezer trap. (a)~The presence of the microwaves forms two dressed states that diverge as a function of the intermolecular separation due to the dipole-dipole interaction. Coupling to the upper state results in a repulsive interaction that prevents the molecules from reaching short range and thus suppressing collisional loss, while coupling to the lower state enhances the loss. (b)~The lifetime of the molecule-pair in the ground state to those prepared in either of the dressed states. \textit{Source: (a, b) are adapted from Ref.~\cite{Anderegg2021}.} }
\label{fig:Doyle_Shielding}
\end{figure}

An alternative approach that is now widely being adopted is microwave shielding~\cite{Karman2018,Lassabliere2018}. In its simplest form, this uses a circularly polarised microwave field to address $\sigma^+$~transitions in the molecule, with a frequency that is blue-detuned from resonance. This field couples pairs of colliding molecules in $N=0$ to a repulsive potential associated with one molecule in $N=0$ and the other in $N=1$. Microwave shielding was first demonstrated for pairs of CaF molecules confined in an optical tweezer, as shown in Fig.~\ref{fig:Doyle_Shielding}~\cite{Anderegg2021}. At a typical molecule temperature of $96\,\upmu$K, the collisions occurred in three dimensions, and the microwaves reduced the rate of loss by a factor of six compared to the bare collision rate. 

Microwave shielding was first applied in bulk gases for fermionic NaK molecules, and was an important component in achieving quantum-degenerate Fermi gases for this species~\cite{Schindewolf2022}. Microwave shielding significantly suppressed the loss rate of these molecules and yielded a ratio of elastic to inelastic collisions~$\ge500$ using a microwave coupling strength of $h\times 11$\,MHz that is blue-detuned from resonance by~8\,MHz. 

The intermolecular potential that results from microwave shielding has an attractive component that can support bound states. The energy of these states with respect to that of free molecules can be tuned by varying the parameters of the microwave field. This gives rise to the presence of field-linked resonances~\cite{Avdeenkov2003,Avdeenkov2004} that allow tuning of the scattering properties of the molecules in much the same way that Feshbach resonances enable tuning of atomic collisions. Chen~\emph{et~al.} identified two field-linked resonances for fermionic NaK, and by tuning the microwave frequency and polarisation demonstrated tuning of the inelastic collision rate by 3 orders of magnitude~\cite{Chen2023}. Following the observation of these resonances, this group also demonstrated how these bound states may be coherently populated to form cold and dense samples of weakly-bound tetratomic (NaK)$_2$ molecules~\cite{Chen2024}. 

Shielding with a single microwave field has been applied to the bosonic molecules NaCs~\cite{Bigagli2023} and NaRb~\cite{Lin2023}. In each case, a significant reduction in the loss rate was observed and efficient evaporative cooling demonstrated. However, in each case Bose-Einstein condensation required a more complex scheme~\cite{Bigagli2024,Shi2025}. The presence of the field-linked bound states leads to enhanced 3-body loss at high densities for bosonic molecules~\cite{Stevenson2024}. The solution to this problem is to introduce a second microwave field, this time linearly polarised along the quantisation axis such that it addresses $\pi$~transitions. This additional field minimises the attractive component of the interaction at long range to realise a purely repulsive potential, thus eliminating the problematic bound states~\cite{Karman2025}. Moreover, using this double microwave shielding scheme affords additional control allowing the strength, anisotropy and length-scale of the interactions to be tuned independently of one another.

\subsection*{Quantum-degenerate gases of polar molecules}

\begin{figure}[t!]
\centering
\includegraphics[width=1.0\columnwidth]{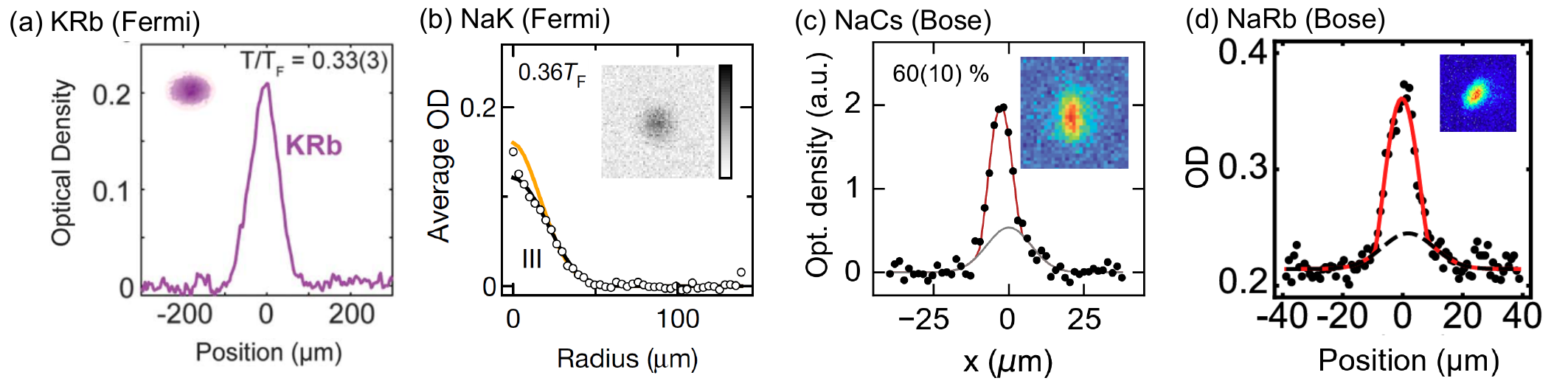}
\caption{Quantum-degenerate Fermi gases of (a)~KRb, (b)~NaK and Bose-Einstein condensates of (c)~NaCs, (d)~NaRb realised in experiments. For each, the optical density extracted from absorption images (shown inset) is plotted as a function of position. For (c,d), there are two lines that show fits to the condensate and thermal fractions of the sample, respectively. \textit{Source: (a) is adapted from Ref.~\cite{DeMarco2019}, (b) is adapted from Ref.~\cite{Schindewolf2022}, (c) is adapted from Ref.~\cite{Bigagli2024}, (d) is adapted from Ref.~\cite{Shi2025}.} }
\label{fig:QuantumGases}
\end{figure}

Bringing ultracold molecules into quantum degeneracy is a major landmark, signifying full quantum control over collisions and molecular gases. The first quantum-degenerate Fermi gas was reported with KRb by De~Marco~\emph{et~al.}~\cite{DeMarco2019}. These experiments started with a very cold mixture containing a sufficiently large number of atoms that the associated atoms remained in the degenerate regime. The samples prepared were characterised by $T/T_F$ ranging from 0.3 to 1, where $T$ is the temperature of the sample and $T_\mathrm{F}$ is the Fermi temperature~\cite{Giorgini2008}. The coldest sample reported is shown in Fig.~\ref{fig:QuantumGases}(a). In a follow up work Tobias~\emph{et~al.} showed that following association, the weakly-bound Feshbach molecules rapidly thermalise with the initial atomic mixture, which is key to how the molecular gas reaches a thermal equilibrium, and this equilibrium is maintained through the ground-state transfer. As an additional confirmation of degeneracy, they observed sub-Poissonian density fluctuations in both the Feshbach and ground-state molecule samples~\cite{Tobias2020}. Quantum-degenerate 2D gases have also been prepared for KRb by loading the atomic mixture into an optical lattice before association. In these experiments, the molecules were initially above the Fermi temperature and then evaporatively cooled to below the Fermi temperature using non-resonant dc field shielding~\cite{Valtolina2020}. The optimum 2D gas had $T/T_\mathrm{F}=0.6(2)$. 

Schindewolf~\emph{et~al.} have prepared a degenerate Fermi gas of NaK molecules~\cite{Schindewolf2022}. This experiment exploited single-frequency microwave shielding to evaporatively cool the molecules below the Fermi temperature, achieving $T/T_F=0.36$, as shown in Fig.~\ref{fig:QuantumGases}(b). More recently, Duda~\emph{et~al.} have shown that with careful matching of the densities of the initial bosonic and fermionic atomic gases, strong boson-fermion interactions induce a phase transition from a polaronic condensate to a molecular Fermi gas. This can be exploited to enhance the efficiency of magnetoassociation enabling the direct production of degenerate Fermi gases of NaK with $T/T_\mathrm{F}=0.52(2)$~\cite{Duda2023}.

The first Bose-Einstein condensate (BEC) of polar molecules was realised by Bigagli~\emph{et~al.} with NaCs~\cite{Bigagli2024}. This experiment utilised the double microwave shielding scheme~\cite{Karman2025} to suppress the loss at high densities that is otherwise present due to 3-body recombination~\cite{Stevenson2024}. They evaporatively cooled the molecules from an initial temperature of 0.70(5)\,$\upmu$K to 6(2)\,nK over 3~seconds. They generated samples containing around 2000 molecules at the onset of BEC, and around 200 molecules for nearly pure samples~\cite{Bigagli2024}, as shown in Fig.~\ref{fig:QuantumGases}(c). The lifetime of the BEC in the presence of the microwave fields was around 2 seconds in an optical dipole trap. Shi~\emph{et~al.} recently applied a similar scheme for NaRb molecules yielding BECs containing approximately 500~molecules with a maximum BEC fraction of~70\%, shown in Fig.~\ref{fig:QuantumGases}(d). 

\begin{figure}[t]
\includegraphics[width=1.0\columnwidth]{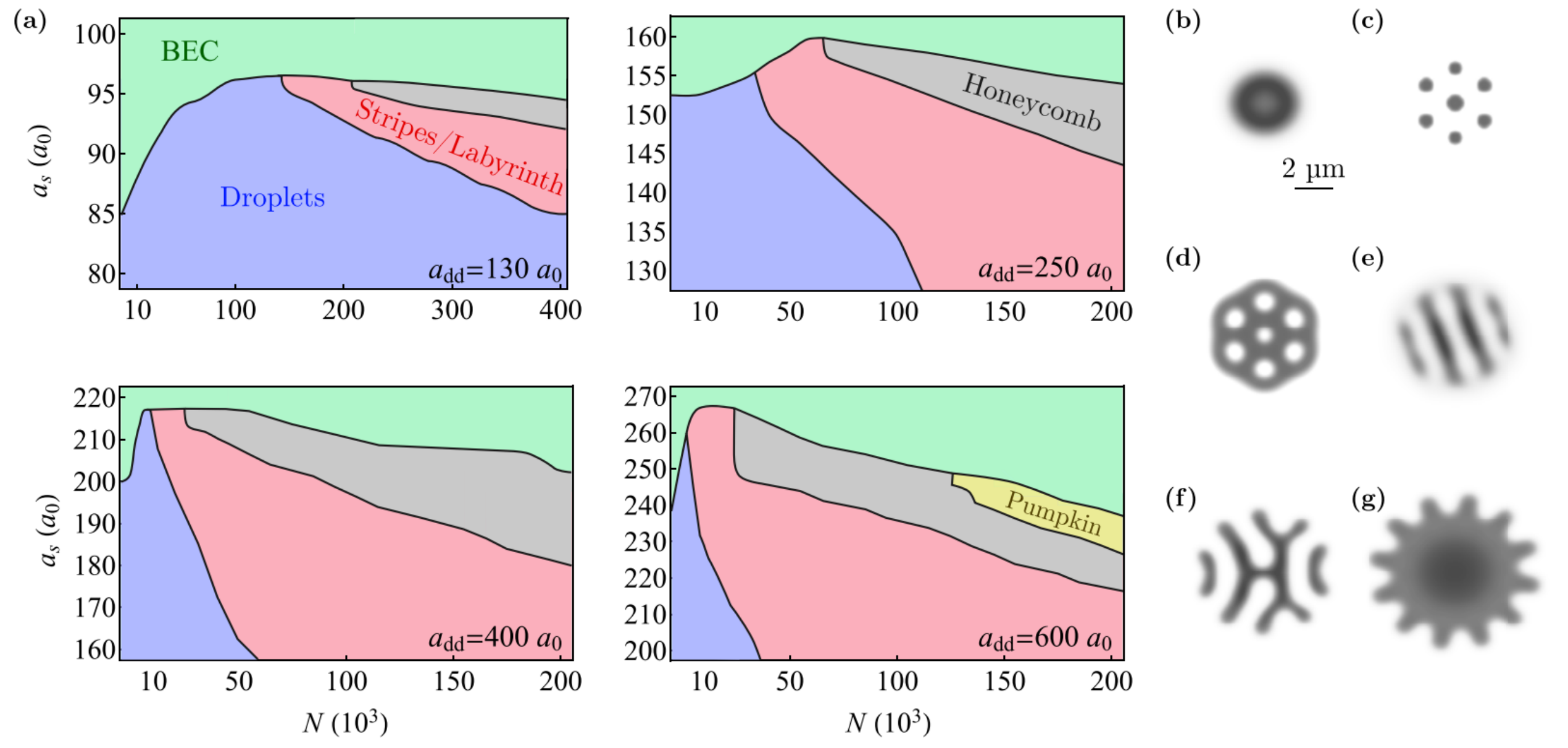}
\caption{Supersolid states of dipolar molecular BEC. (a)~The phase diagram is shown as a function of the $s$-wave scattering length and number of molecules for various values of the dipolar length $a_\mathrm{dd}=d^2m/(12\pi\hbar^2\epsilon_0)$, where $m$ is the mass of the molecule and $\epsilon_0$ is the vacuum dielectric constant. (b-g) show the density distributions associated with these states, labelled as (b)~BEC, (c)~droplets, (d)~honeycomb, (e,f)~stripe/labyrinth, (g)~pumpkin. \textit{Source: figure is adapted from~\cite{Schmidt2022}.} }
\label{fig:droplets}
\end{figure}

The shielding for Bose-Einstein condensation for both NaCs and NaRb is tuned such that the molecules are in a weakly-interacting regime; the intermolecular potential is dominated by a repulsive hardcore van der Waals potential that scales as $1/r^6$, where $r$ is the intermolecular separation. Tuning away from this regime introduces a long-range dipolar term that instead scales as $1/r^3$. Introducing the dipolar term enables the exploration of a rich phase diagram comprised of various discrete supersolid and self-bound droplet phases which have been theoretically predicted~\cite{Schmidt2022,Jin2025,Langen2025}, as shown in Fig.~\ref{fig:droplets}. Very recent theoretical work predicts that droplets should form at interaction strengths lower than previously predicted and, intriguingly, may transition to a superfluid membrane and then a crystalline monolayer as interactions are increased~\cite{Ciardi2025}. Self-bound droplets, similar to those illustrated in Fig.~\ref{fig:droplets}(c), are now beginning to be observed in experiments~\cite{Zhang2025,Shi2025}. Future experimental and theoretical studies are set to reveal more information about the phase diagram associated with these long-lived and highly-dipolar systems. Moreover, many interesting questions remain relating to how the large repulsive shielding core modifies the physics of these gases.

\section{Dipolar spin physics}

Arguably, the most exciting feature of molecules in the context of quantum science is the combination of a vast array of internal rotational and hyperfine states with controllable dipole-dipole interactions. So far, we have mostly focused on samples of molecules prepared in a single internal state.
Now, we will discuss efforts towards quantum simulation and computation using multiple internal states in large scale arrays coupled by dipolar interactions. 
We will review the experiments which have studied many-body physics with molecules, starting with the seminal first observation of dipolar spin-exchange interactions \cite{Yan2013} and follow the progression of the field to the present day, where experiments are able to deterministically entangle individually controlled molecules in optical tweezers.

\begin{figure}[p]
    \centering
    \includegraphics[width=\linewidth]{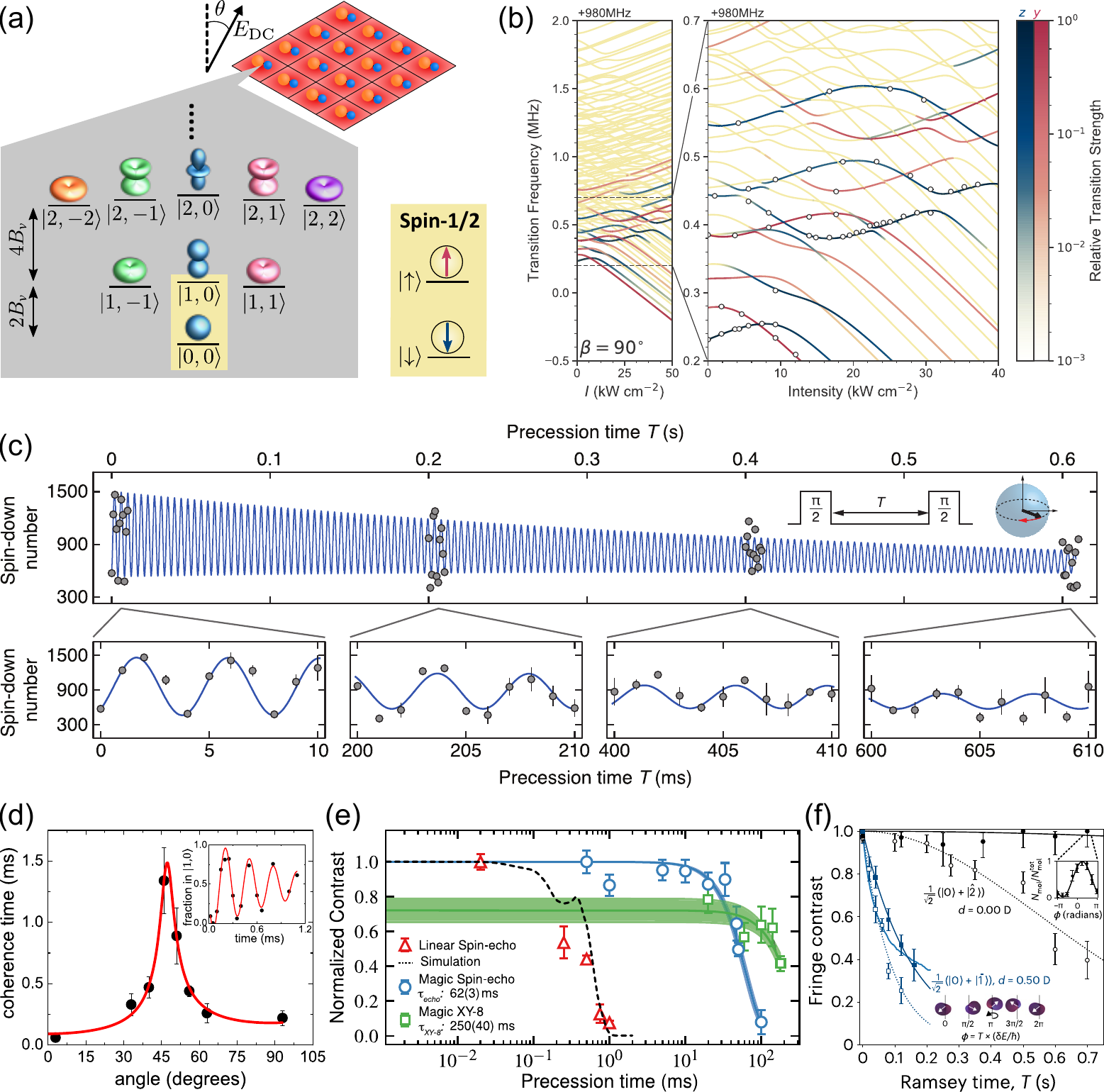}
    \caption{Engineering long-lived coherences between molecular states. (a) Studies of spin dynamics use molecular states, such as rotational levels, to encode pseudo-spins. Molecules are pinned to sites in ordered arrays and experimentalists observe molecule-molecule correlations and spin transport. For each rotational state $N,M_N$ the wavefunction is illustrated. (b) Frequencies of transitions between the ground and first-excited rotational states in RbCs. The molecules are trapped in light that has a wavelength of 1550\,nm, and which is linearly polarised perpendicular to the quantisation axis. Rotational coherence is maximised at the turning points where, to first order, the transition frequency is constant with intensity. (c) Second-scale coherence between hyperfine states of NaK molecules. The electronic wavefunction is decoupled from the nuclear spins, so differential light shifts are minimised. (d) Extended coherence between rotational levels of KRb molecules by setting the polarisation of the trapping light at a magic angle which minimises the differential light shift. The inset shows the Ramsey fringes at the longest coherence time achieved. (e) Rotational coherence extended to the 10\,ms scale for individually-trapped NaCs molecules in optical tweezers by setting the light to a magic ellipticity. (f) Second-scale coherence achieved between rotational state superpositions in RbCs molecules by trapping them in a dipole trap formed from light at a magic wavelength. The lines labelled as $(\ket{0}+\ket{\bar{1}})/\sqrt{2}$ correspond to experiments with dipole-dipole interactions, while the lines labelled as $(\ket{0}+\ket{\hat{2}})/\sqrt{2}$ are non-interacting. In each case, the solid and dotted lines corresponds to Ramsey sequences with and without a spin echo pulse respectively. The inset shows an example Ramsey fringe for the non-interacting case at the longest hold time of 0.7\,s.  \textit{Source: (a) is adapted from Ref.~\cite{Blackmore2019}, (b) is adapted from Ref.~\cite{Gregory2017}, (c) is adapted from Ref.~\cite{Park2017}, (d) is adapted from Ref.~\cite{Neyenhuis2012}, (e) is adapted from Ref.~\cite{Park2023}, (f) is adapted from Ref.~\cite{Gregory2024}.} }
    \label{fig:coherence}
\end{figure}

First, it is helpful to revisit molecular interactions in the context of spin physics.
As introduced in Sec.~\ref{sec:structure:interactions}, molecules in neighbouring rotational states that are connected by an allowed transition support dipolar interactions.
For studies of spin dynamics, one can encode a spin-$1/2$ system using two such states [see Fig.~\ref{fig:coherence}(a)].
Then, in the presence of an applied electric field, Eq.~\eqref{eq:structure:dipolar-hamiltonian} can be recast into the well-known XXZ Hamiltonian \cite{Wall2015} 
\begin{equation} \label{eq:interaction-hamiltonian}
    H_\mathrm{XXZ} = \frac{1}{2} \frac{1-3\cos^2\theta}{r^3}\left[\frac{J_\perp}{2}\left(\sigma^+_1\sigma^-_2 + \sigma^-_1\sigma^+_2\right) + J_z\sigma^z_1\sigma^z_2\right].
\end{equation}
Here, $\sigma^+_i$, $\sigma^-_i$, and $\sigma^z_i$ are the usual spin raising, lowering, and $z$ operators, respectively for molecule $i$, and $\theta$ is the angle between the intermolecular separation and the quantisation axis.
$J_\perp$ quantifies the strength of spin-exchange interactions and $J_z$ the strength of Ising interactions.
The exact values of $J_\perp$ and $J_z$ depend on the rotational states that one chooses and the strength of the applied electric field, but they are proportional to the molecule-frame dipole moment and are on the kHz-scale for $\upmu$m separations (see Table~\ref{Tab:1}).
Notably, $J_z$ vanishes in zero electric field; in this case one recovers the XX Hamiltonian $H_\mathrm{XX} \propto J_\perp \left(\sigma^+_1\sigma^-_2 + \sigma^-_1\sigma^+_2\right) = 2J_\perp \left(\sigma^x_1\sigma^x_2 + \sigma^y_1\sigma^y_2\right).$

In this section, we will first cover the rich physics that has been studied in the XX regime with molecules pinned to the sites of an optical lattice before extending to more recent work implementing the XXZ model.
Then, we will move onto how these interactions can mediate quantum entanglement for quantum computing applications with molecules.

\subsection{Coherences in ultracold molecules}\label{sec:coherences}

For the \emph{resonant} spin-exchange interactions to be harnessed for studies of spin dynamics and wider quantum science, it is essential that one can engineer rotational state superpositions with coherence times much greater than the typical interaction timescales.
Practically, this requires coherence times on the order of $\sim100\,$ms between a chosen pair of molecular states.
In this context, we look at two degrees of freedom commonly used in ultracold molecules: the hyperfine states, which can be used for robust information storage, and the rotational states, which support the dipolar interactions used for studies of spin exchange. 
 
Measurements of coherence are usually performed using Ramsey spectroscopy. Here a $\pi/2$ pulse initialises the system in a superposition of the two coupled states. This is then allowed to freely evolve for a time before another $\pi/2$ pulse projects the accumulated phase onto the populations of the states. Ramsey fringes may be observed as a function of time, frequency, or the phase of the second pulse, and the contrast of the fringe is a measure of the coherence at the time of the final pulse. 

A widely used approach for extending coherence times is to use dynamical decoupling. These are sequences of microwave pulses that were first developed for nuclear magnetic resonance (NMR) spectroscopy, and are now commonly applied for noise suppression and error correction in a range of quantum systems~\cite{Viola1999}. Perhaps the simplest of these sequences is the spin-echo pulse, which simply introduces a $\pi$ pulse half way through the free-evolution time, though more complex sequences utilising many pulses have also been used in recent experiments~\cite{Yan2013,Burchesky2021,Park2023,Langen2024}.

\subsubsection*{Hyperfine coherences}

First, we discuss engineering long-lived coherences in the hyperfine structure of the rotational ground state. These states all have the same rotational angular momentum $N=0, M_N=0$, but differ in the projection of the nuclear spins onto the quantisation axis . For $\mathrm{X}^1\Sigma$ molecules, such as the bialkalis, the energy separation between hyperfine states is generally $<h\times1$\,MHz, as shown in the lower panels of Fig.~\ref{fig:Zeeman}. Crucially, these hyperfine states have properties that make them ideal for robust information storage~\cite{Park2017,Ni2018}. The energy difference between hyperfine states is relatively insensitive to magnetic fields as their magnetic moments are derived from nuclear magnetic moments, and they are highly insensitive to variation in electric and off-resonant optical fields. Moreover, transitions between the hyperfine states are forbidden by selection rules which eliminates problems associated with interaction-induced dephasing.

Long coherence times between $N=0$ hyperfine states were first demonstrated by Park~\textit{et al.}~\cite{Park2017}, who used a gas of fermionic NaK molecules confined in an optical dipole trap.
They achieved a coherence time of $0.7(3)\,$s between the states with nuclear spin projections $(m_\mathrm{Na}=3/2,m_\mathrm{K}=-4)$ and $(m_\mathrm{Na}=3/2,m_\mathrm{K}=-3)$.
Their results are shown in Fig.~\ref{fig:coherence}(c); they performed Ramsey spectroscopy and decoherence manifests in an exponential decay of the fringe contrast as a function of time.
Subsequent work with RbCs~\cite{Gregory2021} realised coherence times beyond $5.6\,$s by carefully tuning the magnetic moments of the two hyperfine states to be identical. This revealed a very weak second-order differential tensor light shift between the states, which the authors were also able to eliminate by tuning the polarisation of the trapping light.
Similar coherence times have been obtained for NaRb molecules trapped in a three-dimensional optical lattice, where collisional losses are also eliminated~\cite{Lin2022}.
These long interrogation times allow measurements of transition frequencies to sub-Hz-level precision.
For example, Park~\textit{et al.}~\cite{Park2017} place an upper bound on collisional interaction shifts to the 100\,mHz level, which was only possible with such long coherence times.

\subsubsection*{Rotational coherences}
Now, we discuss engineering long-lived coherence for rotational-state superpositions that support spin-exchange interactions. This is challenging, as optically trapped molecules generally experience large differential light shifts between rotational states, as discussed in Sec.~\ref{sec:structure:trapping}. Therefore as the molecules sample different trap intensities, their rotational transition frequencies vary. This causes rapid dephasing across the sample, an associated loss of coherence and suppression of spin-exchange interactions.

The interaction between a molecule and off-resonant optical trap light can be seen in Fig.~\ref{fig:coherence}(b), which shows the light shift of transitions from $N=0\rightarrow 1$ in RbCs as a function of the trap intensity~\cite{Gregory2017}. In the presence of hyperfine structure, the tensor polarisability results in a complex structure of avoided crossings with both linear and quadratic shifts in energy that can be tuned with the laser polarisation. Where there are turning points, specific transitions become relatively insensitive to intensity variations in the trap, minimising the dephasing. Blackmore~\textit{et al.} found an enhancement in rotational coherence associated with a particular pair of states when the trap intensity was chosen to be close to such a turning point~\cite{Blackmore2019}, though the maximum coherence time observed was only 0.75(6)\,ms. More recently, this technique has been used with a lattice of NaRb molecules by Christakis \textit{et al.}~\cite{Christakis2023}, who achieved single-molecule coherence times of 56(2)\,ms.

A more general strategy for extending rotational coherence is to align the polarisation of the trapping light at a magic angle.
This was proposed in 2010 by Kotochigova and DeMille~\cite{Kotochigova2010} and can be understood by considering the form of Eq.~\eqref{eq:lightshift}.
By setting $P_2(\cos\theta)=0$ (i.e.\ $\theta\approx55^\circ$), one eliminates the first-order differential polarisability between rotational states.  
This technique was experimentally demonstrated by Neyenhuis \textit{et al.}~\cite{Neyenhuis2012} in a gas of KRb molecules.
Fig.~\ref{fig:coherence}(d) shows their results: by tuning the polarisation of the trapping light, they achieved a coherence time of $1.5(2)\,$ms when driving the transition $(N=0, M_N=0)\rightarrow(1,0)$.
This technique has since been applied in a range of other experiments~\cite{Seesselberg2018,Blackmore2020,Burchesky2021}.
It is particularly effective for molecules with large hyperfine splittings, for example molecules with \(^2\Sigma\) ground states such as CaF. In this situation, the rigid-rotor approximation underlying the magic-angle approach is valid within each hyperfine sub-space. Rotational coherence times of up to $100$\,ms have been observed using this approach for CaF molecules in optical tweezers \cite{Burchesky2021}. 
A related approach, particularly applicable when molecules are confined in high-intensity optical tweezers, was developed by Park~\textit{et al.}~\cite{Park2023} who trapped NaCs molecules in light at a magic ellipticity.
These results are shown in Fig.~\ref{fig:coherence}(e). By incorporating echo pulses, they achieved coherence times of up to $250(40)\,$ms.
Crucially for studies of spin dynamics, these methods enable coherence times much longer than characteristic interaction timescales given in Table~\ref{Tab:1}.

An alternative approach is to use a magic wavelength, that is, to change the wavelength of the trap light such that two rotational states have the same polarisability. So far implementations of magic-wavelength traps for rotational transitions in molecules have been with bialkali molecules, and all exploited the nominally-forbidden transition from $\mathrm{X}^1\Sigma$ to the lowest vibrational states of $\mathrm{b}^3\Pi$. These transitions may only be driven due to mixing with the nearby $\mathrm{A}^1\Sigma$ potential, and are narrow~\cite{Kobayashi2014, Bause2020, He2021, Yang2025, Das2025}. Bause~\textit{et al.} first used this transition to create a magic-wavelength trap for fermionic NaK molecules~\cite{Bause2020}. Their approach however also required the application of an electric field which limited the coherence time to around 1\,ms. 

Theoretical work by Guan~\textit{et al.} showed that for heavy bialkali molecules, such as RbCs, there is a magic condition between the lowest-energy transitions to $\mathrm{b}^3\Pi$ that completely nullifies the anisotropic polarisability $\alpha^{(2)}$ such that all of the tensor light shifts are eliminated~\cite{Guan2021}. This is possible because the $\mathrm{b}^3\Pi$ transitions affect $\alpha_\parallel$ but not $\alpha_\perp$ allowing the setting of $\alpha_\parallel=\alpha_\perp$. This is not possible for lighter molecules, such as NaK, as the condition is only met at a detuning that is small compared to the rotational splitting in the excited state. This type of magic-wavelength trap was demonstrated experimentally using a bulk gas of RbCs molecules~\cite{Gregory2024}. In the absence of interactions, guaranteed by examining the coherence between $N=0$ and~2, a Ramsey coherence time of 0.78(4)\,s was achieved. This was extended to over~$1.4$\,s with a spin-echo sequence; much longer than typical interaction timescales required for studies of spin dynamics. Indeed, when using rotational superpositions that generate dipolar interaction much shorter coherence times were observed, which was limited by interaction-induced dephasing~\cite{Gregory2024}. Crucially, the detuning of the magic wavelength from the transition was 186\,GHz, much larger than the $\sim14$\,kHz linewidth associated with the nearest transition~\cite{Das2025} such that the trap is compatible with long lifetimes. This magic-wavelength technique has since been extended to individually trapped RbCs molecules, where a rotational coherence time of 15(6)\,s was achieved without spin echo~\cite{Ruttley2025}. 
Further, this technique has been simultaneously applied to 3~transitions, enabling the coherent encoding of spin-1 dynamics in the rotational levels of RbCs~\cite{Hepworth2025}.

These advances demonstrate the remarkable degree of control now achievable over molecular coherence.
We now turn our attention to experiments which exploit this level of control for studies of dipolar spin physics.

\subsection{Coherent interactions of molecules trapped in optical lattices}
In this section we consider dipolar interactions studied in molecules trapped in optical lattices, periodic potentials formed by standing waves of light. This technique can be used to create large arrays of nearly-uniform optical traps, with typical array spacings between 500~nm and 10~$\upmu$m. We follow the progress of the field from the initial observation of dipolar spin-exchange interactions to present day experiments, which are able to resolve correlations at the single-molecule level. 

Molecules in optical lattices are well suited to the simulation of condensed matter physics, where the periodic structure of the lattice enables a direct comparison with models used to study periodic solids. In a pioneering proposal Micheli~\emph{et al.} showed how molecules could be used to engineer lattice spin models with topologically protected ground states \cite{micheliToolboxLatticespinModels2006}.
This was further developed by a series of works exploring quantum magnetism \cite{gorshkovQuantumMagnetismPolar2011,gorshkovTunableSuperfluidityQuantum2011}, exotic superfluidity \cite{kuns$d$waveSuperfluidityOptical2011} and dipolar extended Hubbard models \cite{wall_emergent_2009,wall_hyperfine_2010,wall_molecular_2013}. All of these proposals strongly motivated the formation of a gas of polar molecules in the lowest band of an optical lattice, and the development of full control of the dipolar interaction using external electric and microwave fields.

\subsubsection*{Low entropy molecular arrays in lattices}

The first experiments with ultracold molecules in optical lattices were performed with KRb. Chotia~\textit{et al.}~\cite{Chotia2012} demonstrated that loading the molecules into a 3D lattice was a highly-effective method of stopping the collisional loss discussed in Section~\ref{sec:3}. By pinning the molecules in place, they realised the first collisionally stable gas of polar molecules with a lifetime of up to 25\,s. The introduction of the optical lattice also lends the advantage that interactions can only occur at quantised strengths associated with the separation between lattice sites. 

\begin{figure}[t]
    \centering
    \includegraphics[width=\linewidth]{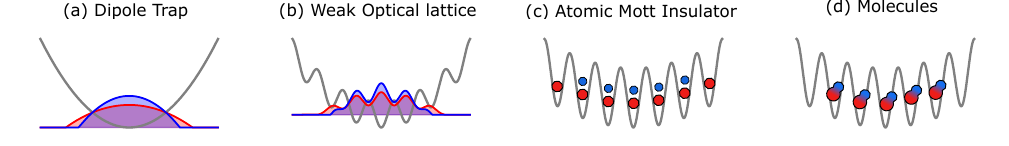}
    \caption{Overview of the process for the production of low entropy arrays of molecules using association in optical lattices as proposed in \cite{jakschCreationMolecularCondensate2002a} and realised in \cite{Moses2015,Reichsoellner2017}. (a)~A quantum-degenerate mixture of two atomic species is prepared in an optical dipole trap. (b)~This mixture is loaded adiabatically into an optical lattice, such that all the atoms occupy the lowest band of this optical lattice. (c)~The lattice depth is increased until the system transitions to a Mott insulating state with one atom of each species per site. (d)~The atom pairs are associated to form single molecules occupying each site of the lattice. Note that for both atoms occupying the ground motional state of the lattice, there is maximal phase space overlap between the two atoms such that the efficiency of association approaches unity.}
    \label{fig:making molecules in lattices}
\end{figure}

A highly-filled, or even defect-free, array is the ideal starting point for lattice experiments. An approach for these initial conditions was first identified by Jaksch~\emph{et~al.} who proposed creating a molecular BEC from the association of pairs of atoms in a Mott insulator state with exactly two atoms per site~\cite{jakschCreationMolecularCondensate2002a}; this was shortly after the first atomic experiments demonstrated the quantum phase transition from a BEC to a Mott insulator~\cite{Greiner2002}. This idea was soon extended to show that a state with one of each species could be used to form a condensate of \emph{polar} molecules \cite{damski_creation_2003,mooreControllingTwospeciesMottinsulator2003a}, as illustrated in Fig.~\ref{fig:making molecules in lattices}.

This approach was demonstrated to work very effectively for homonuclear molecules in \cite{danzlUltracoldHighdensitySample2010a}, where a sample of ground-state Cs\(_2\) molecules were prepared in the ground band of a 3D lattice with approximately 85\% filling. However, in practice, the dual-species Mott insulator state is significantly more difficult to engineer than the single-species counterpart. In the case of K-Rb mixtures it was found that the differing quantum statistics of the atoms required the creation of a highly imbalanced mixture with very small Rb condensates to prevent multiply occupied sites~\cite{Moses2015}. The peak lattice-filling fraction reported in these experiments was on the order of 30\%. 
Further investigations of the physics of the Bose-Fermi mixture in the lattice elucidated some of the doublon physics which complicates the association process~\cite{Covey2016}.

The dual-species Mott insulator approach was also demonstrated for Rb-Cs Bose-Bose mixtures. In this case, difficulty arises from a large interspecies background scattering length which leads to strong three-body losses and makes the mixture immiscible~\cite{McCarron2011}. To better control the interspecies collisions, an approach was developed where the species were cooled separately and then merged in an optical lattice after first forming a Mott insulator of Cs~\cite{Reichsoellner2017}. This allowed the interspecies interactions to be tuned to zero during the merging using an interspecies Feshbach resonance. As in KRb, the maximum filling fraction reported was around 30\% for weakly-bound Feshbach molecules. Later work established a new STIRAP route for RbCs molecules formed using this approach~\cite{Das2023}. Dual-species Mott insulators have also been used to form NaRb molecules; in this case realising a filling of around 15\% \cite{Rosenberg2022}. 

With the advent of collisional shielding techniques and the preparation of degenerate gases, it may soon be possible to instead directly load a degenerate gas of molecules into the lattice to form a Mott insulator of molecules with one molecule per site. However, this approach will require careful control of the dipolar interactions, as well as sufficiently-high condensate fractions, to realise low-entropy lattice filling.

\subsubsection*{First observation of dipolar spin exchange with polar molecules}

\begin{figure}[t]
    \centering
    \includegraphics[width=\linewidth]{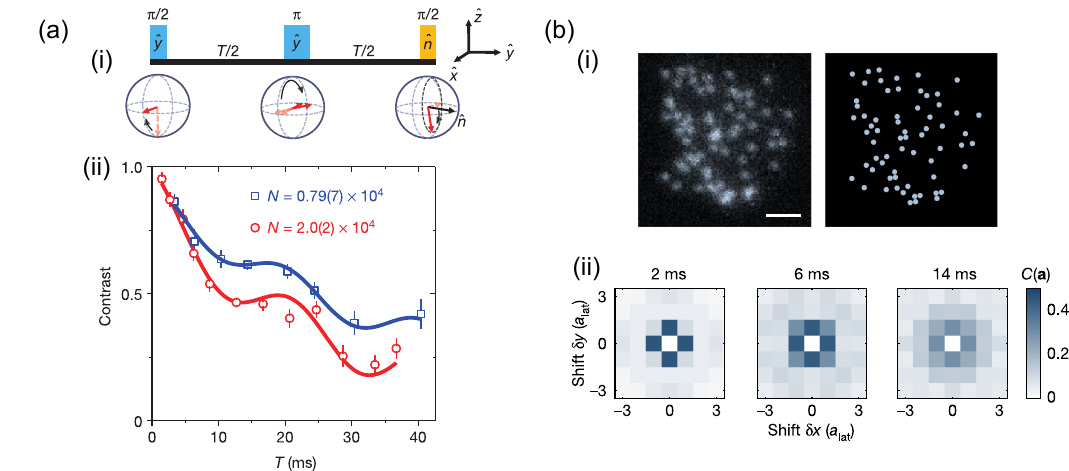}
    \caption{
    Probing interactions between ultracold molecules in optical lattices. (a) First observation of coherent dipole-dipole interactions with KRb molecules. (i) Overview of the spin-echo sequence used to prepare and probe an interacting superposition of rotational states. A Bloch sphere representation of the state evolution is shown below the pulse diagram. (ii)  Entanglement between molecules results in variation of the contrast in Ramsey fringes; they observe oscillations and decay of the Ramsey fringes associated with dipolar spin-exchange interactions in the lattice. 
    (b)~Detection of individual NaRb molecules in an optical lattice using a quantum gas microscope. (i)~The molecules are dissociated and their locations tagged by a single Rb atom that is detected by fluorescence imaging. The scale bar in the image represents 5$\mu$m, corresponding to around 7 lattice sites. (ii)~Combining the spin-echo sequence with quantum gas microscopy allows resolution of the spin-spin correlations induced by the dipole-dipole interactions \cite{Christakis2023}. \textit{Source: (a) is adapted from Ref.~\cite{Yan2013}, (b) is adapted from Ref.~\cite{Christakis2023}.}}
    \label{fig:lattices}
\end{figure}

Following on from the initial observations that optical lattices could be used to suppress the collisional loss of KRb, Yan~\emph{et al.}~\cite{Yan2013} loaded a degenerate mixture of K and Rb atoms into a 3D lattice to create an array of molecules, with filling estimated to be around 10\%, and negligible population of molecules in excited motional states. The combination of the lattice spacing and dipole moment of KRb gave this experiment a nearest-neighbour dipole-dipole exchange energy of $J_\perp\simeq52$\,Hz. Through careful consideration of the effect of the optical lattice light on the internal states of the molecules, they were able to reduce the variation in light shift across the sample to around 500\,Hz, and the site-to-site variations to 6~Hz. Using a spin-echo Ramsey sequence they observed oscillations in the Ramsey fringe contrast, as shown in Fig.~\ref{fig:lattices}(a). These oscillations exhibit the same frequency as that associated with nearest-neighbour spin-exchange interactions. Moreover, the coherence time after spin echo was found to decrease with increasing molecular density, providing further evidence of the many-body dipolar interactions in the sample. As a further test, the experiments employed a Waugh-Huber-Haeberlen (WAHUHA) pulse~\cite{WAHUHA} which reversed the entanglement generated by dipole-dipole interactions between pairs of molecules, but had no effect on higher order interactions involving three or more molecules. This removed the oscillations in the Ramsey contrast, thereby confirming their origin as pairwise interactions.

In a follow up work, Hazzard~\textit{et al.} developed a moving-average cluster expansion (MACE) method to describe the non-equilibrium dynamics of the lattice-confined molecules~\cite{Hazzard2014}. They demonstrate that the system is well-described by an $\mathrm{XY}$~spin model and, by comparing to additional experimental measurements, that the interactions were indeed long-ranged, requiring more than nearest-neighbour or even next-nearest neighbour interactions to accurately reproduce experimental observations.

The demonstration of dipolar spin-exchange interactions spurred on developments to understand the theory of molecules in lattices, particularly in the easier to realise experimental case of finite filling fraction. Hazzard \emph{et al.} explored the dynamics of disordered many-body spin systems after the interaction quench realised in the Ramsey sequence \cite{hazzardFarEquilibriumQuantum2013}, finding that many of the interesting properties persisted at the experimentally realised filling fractions. Other groups investigated applying lattice-confined molecular samples to study other hot topics in quantum condensed matter such as many-body localisation \cite{yaoManyBodyLocalizationDipolar2014}, chiral excitations \cite{syzranovSpinorbitalDynamicsSystem2014,syzranov_emergent_2016}, spin liquids \cite{yao_quantum_2018} and topological insulators \cite{yao_topological_2012,schusterFloquetEngineeringUltracold2021a}. This in turn stimulated the experimental groups to achieve greater lattice filling, aiming to reach the percolation threshold, to explore a greater range of molecules and to improve the detection and control methods in these experiments.

\subsubsection*{Quantum gas microscopy}

Precise detection and control of the position and state of individual molecules in the array is important for the study of correlations, and for the detection of coherent many-body effects. In experiments with atomic gases in optical lattices, the technique of quantum gas microscopy has been revolutionary and has enabled the resolution of quantum correlations between individual atoms in coherent many-body systems~\cite{grossQuantumGasMicroscopy2021}. Applying these techniques to molecules, where the rich internal structure and long-range dipolar interactions allow for a wider range of Hamiltonians to be studied, is an exciting goal for many experiments in the field. 

The first microscope for molecules was reported by the Bakr group~\cite{Rosenberg2022,Christakis2023}. Here, NaRb molecules were prepared in a single plane of a 3D optical lattice. For detection, each molecule is deterministically mapped onto a Rb atom via dissociation and removal of the Na atom, such that each Rb atom tags the locations that were occupied by molecules in the lattice. The Rb atoms are then detected via fluorescence imaging, with a sufficiently high resolution such that the occupied sites of the lattice can be identified with high fidelity.
Exemplary images from this experiment are shown in Fig.~\ref{fig:lattices}(b)(i). The power of the microscopy technique is in allowing direct observation of correlations between particles. The Bakr group first demonstrated this for weakly-bound Feshbach molecules by observing correlations from bosonic quantum statistics~\cite{Rosenberg2022}. This experiment also showed that they had successfully prepared the molecules in the ground band of the optical lattice. Later experiments with ground-state molecules allowed the observation of pair correlations arising from the dipolar spin-exchange interaction \cite{Christakis2023}, as shown in Figure \ref{fig:lattices}(b)(ii).

Covey \emph{et al.} \cite{Covey2018} proposed a method to detect both the density and spin of a molecular gas simultaneously via a dual-species quantum gas microscope in which the state of the molecule is mapped onto the atomic species following dissociation. Their theoretical work showed how this would be transformative for studies of disordered spin systems. Recently, this proposal has been realised for RbCs molecules in a dual-species  microscope \cite{mortlockMultistateDetectionSpatial2025}. This work also demonstrated spatially-resolved addressing of the molecular cloud, a key technique for preparing out of equilibrium states for quantum simulation. Many other experiments \cite{gempelAdaptableTwolensHighresolution2019,yaoMeasuringPairCorrelations2025} are being developed for quantum gas microscopy of molecules, motivated by the wealth of opportunities to study many-body physics in a highly tunable and controlled system.

\subsubsection*{Going beyond spin-exchange interactions}

Dipole-dipole interactions between molecules can also be engineered by applying an external electric field to generate static lab-frame dipoles. In the language of effective spin models this leads to an Ising-type term, $J_z\sigma^z_i\sigma^z_j$, realising the more general XXZ model given in Eq.\,\eqref{eq:interaction-hamiltonian}. The Ising term competes with the exchange interaction, and in molecules the relative strength of these two terms can be tuned through the strength of the DC electric field. This allows molecular systems to, for example, reach the symmetric Heisenberg point. The typical electric fields required in experiments are given in Table~\ref{Tab:1} for various molecules. The use of electric fields to control coherent dynamics in experiments can be technically challenging, as it requires exceptionally low-noise, high-voltage electronics and sophisticated pulse sequences for dynamical decoupling. Despite these challenges, the spin dynamics of KRb molecules with tunable Ising interactions, initially investigating the dynamics of collective spins in quasi-2D dipole traps, have been studied~\cite{Li2023}.

An alternative approach to realising a wider variety of spin models without applying DC electric fields is to employ Floquet engineering. Here, periodic driving is used to modify the symmetries and dynamics of a quantum system, and it has been shown that complex spin models can be realised using the resonant exchange interaction alone. For example, in ultracold molecules sudden periodic rotations of the spin basis via $\pi/2$ pulses can be used to mix together spin interactions of the form \(\sigma^i\sigma^i, i\in(X,Y,Z)\). This was first shown in the NaRb microscope experiments~\cite{Christakis2023} and later generalised further in the work of Miller~\emph{et al.}~\cite{millerTwoaxisTwistingUsing2024a} to study models with further reduced symmetry.

A common theme in experiments studying many-body systems with more complex interactions is to identify how to harness the entanglement that emerges for useful metrological gain, for example, by generating spin-squeezed states~\cite{Li2023,millerTwoaxisTwistingUsing2024a}. Such states may be used to perform measurements with precision below the standard quantum limit. This represents an elegant convergence of the modern applications of ultracold molecules in quantum simulation with the roots of the field in precision spectroscopy.

So far we have mainly considered dynamics involving the internal degree of freedom of the molecules only. Allowing the molecules to move around unlocks a much richer platform for quantum simulation. Itinerant molecules can be used to study generalised $t\mbox{-}J$ models, which have long been proposed to explain unconventional magnetic ordering and high-temperature superconductivity. In contrast to atomic systems, in  dipolar molecules the interactions and tunnelling can be tuned independently, which could allow observation of phase transitions at experimentally achievable temperatures. 

Recently, experiments with KRb reported by Carroll \emph{et al.} \cite{Carroll2025} have begun to probe itinerant molecules in lattices, demonstrating independent control over dipolar Ising, spin-exchange and tunnelling degrees of freedom. They studied the interplay of these interactions using Ramsey spectroscopy. There have also been renewed theoretical efforts to model itinerant molecular spin dynamics in finite temperature systems without a lattice \cite{wangTheoryItinerantCollisional2025}. It should be noted that the ability to realise strong electric dipole-dipole interactions while maintaining the long lifetimes required for coherent motion of the particles is unique to molecules among current ultracold platforms, and will certainly be an important direction in future research. These studies of spin dynamics in mobile molecules complement recent developments in complete control of molecular motional states using optical tweezer arrays, as we will consider next.

\subsection{Quantum computing with arrays of molecules}

The rich internal structure and controllable interactions of polar molecules also make them attractive candidates for the building blocks of quantum computers~\cite{Cornish2024}.
DeMille first proposed this in a seminal paper in 2002~\cite{DeMille2002}. 
In the proposed scheme, molecules would be individually trapped in a 1D array, with the quantisation axis provided by an external electric field.
Quantum information would be encoded in the ground and first-excited rotational states, allowing for single-qubit operations with microwave pulses.
Dipole-dipole interactions would be used to mediate two-qubit gates.
In this section, we will cover the progress made in the last decade towards realising this goal. 

As DeMille realised, to harness molecules for quantum computing, one must be able to individually trap, manipulate, and readout molecules with high fidelity.
The technique that has emerged as the preferred choice is to trap and control molecules with optical tweezers: tightly focussed dipole traps with $\upmu$m-scale waists.
These traps are so small that one can engineer ``collisional blockade'' where rapid pairwise loss results in stochastic single-particle loading~\cite{Schlosser2001}.
This method has been used for decades to trap and control individual neutral atoms~\cite{Kaufman2021}, and has now been extended to molecules.
We will start by discussing how single molecules are prepared in arrays of optical tweezers. Then, we will explore how tweezers can be used to control and deterministically entangle molecules, paving the way for quantum information processing and precision metrology in this setting.

 \subsubsection*{Preparing molecules in optical tweezers}
\begin{figure}[t]
    \centering
    \includegraphics[width=\linewidth]{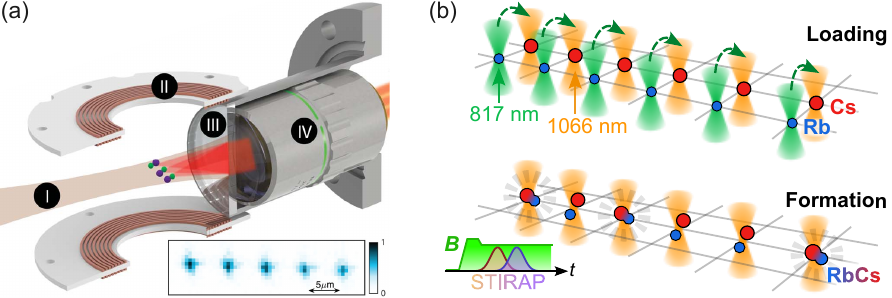}
    \caption{Preparation of molecules in arrays of optical tweezers. (a) The experimental apparatus for a tweezer array of CaF molecules. Single molecules are loaded from an optical dipole trap (I) into tweezers at the centre of the MOT coils (II). A re-entrant vacuum viewport (III) allows the insertion of a high numerical aperture lens (IV) which is used to focus light from an acousto-optic deflector to form the array of tweezers. The same lens is used to image the molecules with fluorescence imaging (see inset). (b) Scheme for assembling molecules (here, RbCs) in optical tweezers. 1D arrays of individual atoms are loaded. Then, the arrays are brought together to prepare atom pairs which are associated into molecules and transferred to the ground state. The assembly efficiency is approximately half. \textit{Source: (a) is adapted from Ref.~\cite{Anderegg2019}, (b) is adapted from Ref.~\cite{Ruttley2024control}.} }
    \label{fig:tweezers-loading}
\end{figure}

First, we will discuss how molecules can be prepared in arrays of optical tweezers.
A typical apparatus for such an experiment is shown in Fig.~\ref{fig:tweezers-loading}(a).
The broad idea is to use a high-numerical-aperture objective lens to form the tightly-focussed optical tweezers.
Molecules are trapped in these tweezers, and the same objective lens can be used to detect them with fluorescence imaging. 

Within the last decade, two complementary approaches have been developed to load these tweezer arrays with individually trapped ultracold molecules . 
In the first approach, one begins with individually-trapped atoms.
Then, these atoms are coherently assembled into molecules. In the second approach, laser-cooled molecules are loaded directly into the tweezers from magneto-optical traps.\\

\noindent\textit{Assembled molecules}\\
The process for assembling molecules in optical tweezers is similar to the methods used with bulk gases and in optical lattices. 
Broadly, one prepares two atoms in a common tweezer trap then associates them into a molecule~\cite{Liu2017}.
Pioneering experiments at Harvard in 2018 by Liu~\emph{et al.}~\cite{Liu2018} first demonstrated this approach with Na and Cs atoms.
Single Na and Cs atoms were trapped in species-specific optical tweezers.
The Cs tweezer was then moved to overlap with the Na tweezer and subsequently switched off, leaving a pair of atoms in a single trap.
The atom pair was photoassociated to form an excited NaCs molecule, detected via atom-pair loss.

Rapid progress followed in improving efficiency and control.
Liu~\emph{et al.}~\cite{Liu2019} performed dual-species Raman sideband cooling on Na and Cs, preparing most atom pairs in the ground state of relative motion to maximise wavefunction overlap; this is crucial for efficient association of atoms into molecules.  
The atom pairs were then transferred to weakly-bound molecular states using a Raman transition.
Later work showed that this association process could be made coherent by carefully selecting the intermediate state used in the Raman process~\cite{Yu2021}.
Alternatively, weakly-bound molecules can be created via magnetoassociation using Feshbach resonances, as demonstrated in optical tweezers with NaCs~\cite{Zhang2020} and RbCs~\cite{Ruttley2023}.

The optical tweezers themselves can also be used to associate atoms into molecules.
Due to the tight confinement of the tweezers, avoided crossings can be engineered between atomic and weakly-bound molecular states as a function of tweezer separation.
Then, as two tweezers are brought together, the trapped atoms are mergoassociated into weakly bound molecules.
This technique has been used to form RbCs molecules~\cite{Ruttley2023} and is predicted to be useful for atomic species with Feshbach resonances that are too narrow for efficient magnetoassociation~\cite{Bird2023,Bird2025}.

As a consequence of the coherent assembly process, all molecules are prepared in a single internal state.
Typically, as in bulk-gas experiments, the weakly-bound molecules are transferred to the absolute ground state with a two-photon process.
Cairncross~\emph{et al.}~\cite{Cairncross2021} first demonstrated this with NaCs molecules in tweezers using a detuned Raman transfer.
Alternatively, STIRAP can be used for high-fidelity transfer~\cite{Guttridge2023}: one-way transfer efficiencies up to $98.7(1)\%$ have been reported for RbCs molecules trapped in optical tweezers~\cite{Maddox2024}.
These techniques are now routinely used to prepare small chains of assembled molecules; a cartoon summarising this assembly procedure is shown in Fig.~\ref{fig:tweezers-loading}(b).

The main limit to the efficiency of molecular assembly in these arrays is the finite temperature of the atoms at the point of association.
When the atoms are motionally excited, it is not possible to adiabatically transfer them into a weakly-bound molecular state.
This motional excitation can be present due to imperfect Raman sideband cooling or heating effects as the atom pairs are brought together.
These typically limit the occupation of the motional ground state to approximately $55\%$~\cite{Zhang2020,Ruttley2024control}, so around half of atom pairs are converted into molecules.
However, this effect means that formed molecules are very cold as only the coldest atoms are associated, and the molecules inherit the centre-of-mass motion of the atom pair.
For example, in arrays of assembled NaCs molecules, $65(3)\%$ of the molecules occupy the 3D motional ground state~\cite{Zhang2022}.\\

\noindent\textit{Laser-coolable molecules}\\
Instead of assembling molecules from ultracold atoms, experiments have demonstrated successful loading of tweezers with molecules that can be directly laser cooled.
The experiments begin with these molecules cooled and loaded into a magneto-optical trap (MOT). Then, they are transferred to the tweezers via an intermediate optical dipole trap.
This intermediate step increases the molecular density to enable efficient loading into the tightly-focused tweezers.
Typically, molecules are transferred from the MOT to the dipole trap while undergoing $\Lambda$-enhanced gray-molasses cooling~\cite{Cheuk2018}.
This cooling reduces the temperature to tens of $\upmu$K while increasing density.
The optical tweezers are then switched on with trap depths comparable to this temperature.
Finally, the larger dipole trap is ramped off to transfer the molecules into the tweezers.
During this process, the cooling light remains on.
As in the case of atomic loading, this light induces rapid pair loss: each tweezer then is either empty or contains a single molecule.
This method was first demonstrated by Anderegg~\emph{et al.}~\cite{Anderegg2019} for CaF molecules.
It was later extended to polyatomic CaOH molecules by Vilas~\emph{et al.}~\cite{Vilas2024}.
Typical loading efficiencies range from $20$--$40\%$~\cite{Anderegg2019,Vilas2024,Holland2023,Bao2023}, primarily limited by the initial molecular densities.

Once loaded into tweezers, the molecules typically occupy a rotationally-excited manifold within the ground electronic potential.
From here, they can be optically pumped to a single hyperfine level in less than $100\,\upmu$s~\cite{Williams2018}.
Then, one can use microwave pulses to coherently transfer them to the absolute ground state.
Typical state-preparation fidelities using these methods are approximately $80\%$~\cite{Holland2023,Bao2023}.

Immediately after loading into tweezers, laser-cooled molecules are significantly hotter than assembled molecules. However, recent experiments with CaF have shown that Raman sideband cooling can be used to cool the molecules to the motional ground state after they have been loaded.
To date, CaF molecules have been cooled to the 3D motional ground state with a probability of up to $54(18)\%$~\cite{Bao2024,Lu2024}, comparable to assembled molecules.

\subsubsection*{Readout and manipulation of individual molecules}
Once molecules are loaded into optical tweezers, they can be individually detected and controlled.  
Detection is typically performed using fluorescence imaging.  
Laser-coolable molecules have a distinct advantage in this regard: they can repeatedly scatter photons, allowing direct imaging with high fidelity.  
The inset of Fig.~\ref{fig:tweezers-loading}(a) shows a typical fluorescence image of CaF molecules in tweezers spaced by approximately $5\,\upmu$m.
In contrast, assembled molecules are usually dissociated into their constituent atoms, which are then imaged.

Spin-sensitive readout can be achieved with both classes of molecules.  
For assembled molecules, different internal molecular states can be mapped onto distinct spatial configurations of the constituent atoms following dissociation~\cite{Ruttley2024control,Picard2024}.
For laser-coolable molecules, spin-specific readout is performed by optically shelving one spin state into a dark state during imaging~\cite{Holland2025}.
Readout infidelities are generally at the percent level, arising from insufficient photon scattering before molecule loss (for laser-coolable) molecules~\cite{Holland2023imaging,Holland2023}, or imperfect state transfer during molecular disassembly (for assembled molecules)~\cite{Maddox2024}.  
These low error rates enable quantum science experiments on small molecular arrays using postselection.

Beyond detection, optical tweezers provide precise control over the position and internal state of individual molecules.  
This enables active correction of array errors by identifying empty sites and rearranging molecules to fill them.  
Rearrangement techniques, originally developed for atomic tweezer arrays~\cite{Miroshnychenko2006Rearrangement,Lee2016Rearrangement,Barredo2016,Endres2016}, were first demonstrated with molecules by Holland~\emph{et al.}~\cite{Holland2023} using CaF, and later extended to RbCs~\cite{Ruttley2024control} and NaCs~\cite{Picard2024} where occupancy was inferred by the detection of atoms on sites where the molecule formation was unsuccessful.
In addition to correcting for empty sites, these techniques can also be used to mitigate state-preparation errors in laser-coolable molecules~\cite{Holland2025}.

Optical tweezers also allow coherent manipulation of molecular internal states.  
For example, auxiliary tweezers can light-shift selected molecules, enabling site-selective rotational excitation with microwave pulses~\cite{Ruttley2024control,Picard2024}.  
This capability has been used to prepare arbitrary spin chains, essential for studying quantum magnetism~\cite{Wall2015} in small molecular arrays or performing single-qubit operations on specific molecules~\cite{Ni2018}.

\subsubsection*{Pair entanglement and multi-qubit gates}
To engineer two-qubit gates with molecules, it is essential that there is a mechanism for quantum-information transfer.
As DeMille proposed~\cite{DeMille2002}, the natural interactions to use to engineer this connectivity are the spin-exchange interactions seen in Eq.~\eqref{eq:interaction-hamiltonian}.
These interactions couple rotational states, and theoretical work by Ni \textit{et al.}~\cite{Ni2018} showed how this interaction could be mapped to an iSWAP gate between hyperfine qubits.\footnote{An iSWAP gate is a two-qubit quantum gate that swaps the states $\ket{\uparrow\downarrow}$ and $\ket{\downarrow\uparrow}$ while adding a phase factor of $i$, leaving $\ket{\downarrow\downarrow}$ and $\ket{\uparrow\uparrow}$ unchanged. When combined with single-qubit gates, it forms a universal gate set for quantum computation.}
Here, we discuss recent experimental work that has implemented these proposals with individually trapped molecules.

\begin{figure}[t]
    \centering
    \includegraphics[width=\linewidth]{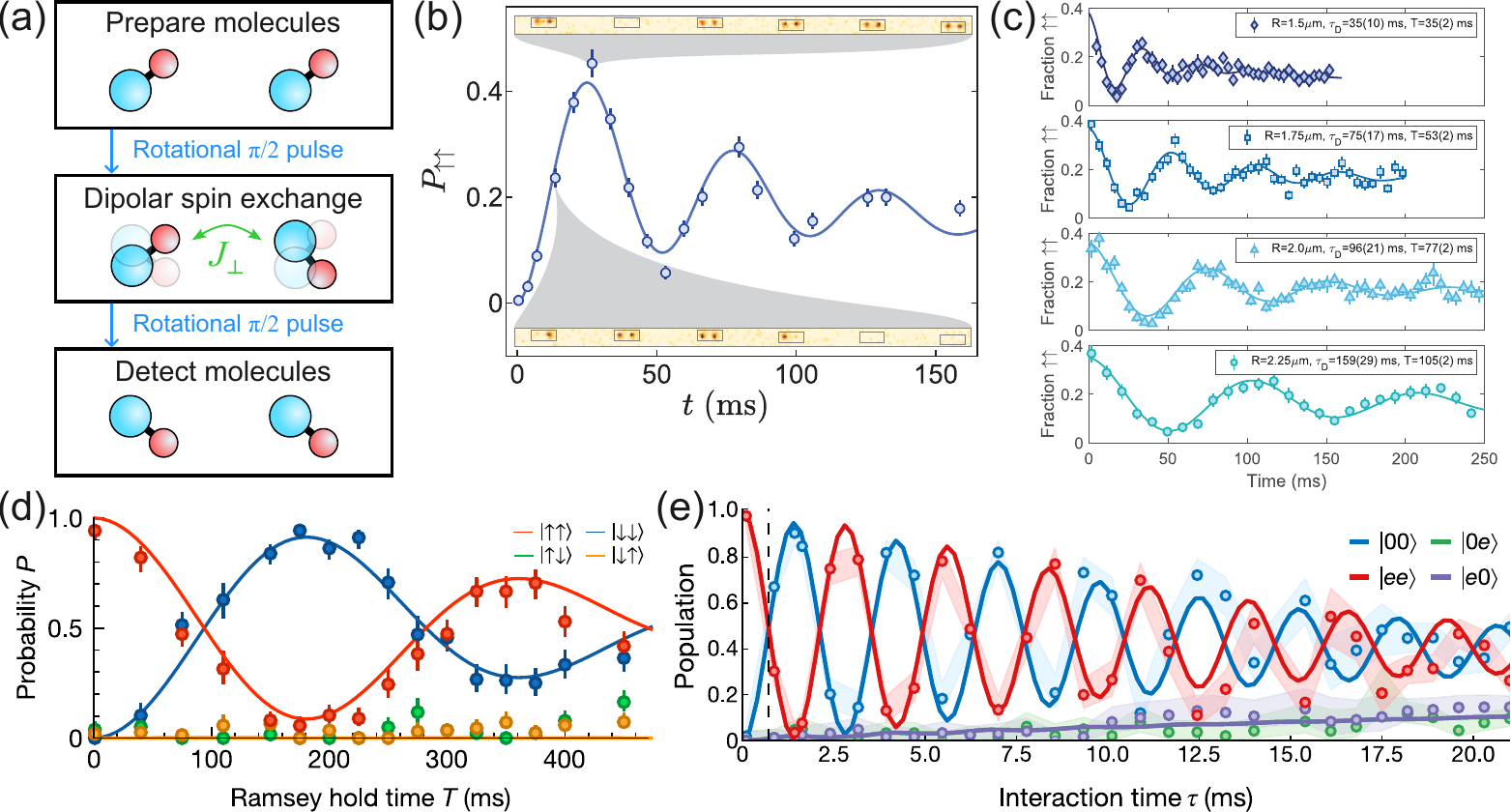}
    \caption{Dipolar spin exchange between pairs of molecules. (a) Experimental sequence for observing spin-exchange interactions, as described in the text. The remaining panels show experimental observations of dipolar spin exchange between (b,c) CaF, (d) RbCs, and (e) NaCs molecules. Each panel shows molecular state populations as a function of interaction time: the oscillations in populations indicate spin-exchange dynamics (see text for details). All molecules are individually trapped in optical tweezers at $\upmu$m-scale separations. \textit{Source: (b) is adapted from Ref.~\cite{Holland2023}, (c) is adapted from Ref.~\cite{Bao2023}, (d) is adapted from Ref.~\cite{Ruttley2025}, (e) is adapted from Ref.~\cite{Picard2025}.} }
    \label{fig:tweezers-entanglement}
\end{figure}

Figure~\ref{fig:tweezers-entanglement}(a) shows the broad experimental procedure used to generate molecular entanglement \cite{Holland2023,Bao2023,Picard2025,Ruttley2025}.
A pair of rotational states $\{\ket{\downarrow},\ket{\uparrow}\}$ that support spin-exchange interactions is chosen.
The molecules are prepared and initialised in a single state, say $\ket{\downarrow}$.
Then, the transition $\ket{\downarrow}\rightarrow\ket{\uparrow}$ is driven with a global $\pi/2$ pulse to put each molecule in an equal rotational-state superposition.
This leads to spin-exchange interactions between the molecules with characteristic strength $J_\perp$.
Finally, after a hold time $T$, a second $\pi/2$ pulse is applied to the rotational transition.
This sequence results in oscillations between pair states $\ket{\downarrow\downarrow}$ and $\ket{\uparrow\uparrow}$ as a function of the hold time with frequency $J_\perp/2$.
To generate a maximally entangled Bell state (i.e.\ an equal superposition of $\ket{\downarrow\downarrow}$ and $\ket{\uparrow\uparrow}$), one allows the molecules to interact for time $T=1/(2J_\perp)$.

In 2023, pioneering experiments at Princeton~\cite{Holland2023} and Harvard~\cite{Bao2023} demonstrated this entanglement routine with pairs of CaF molecules.
The molecules were prepared in tweezers with separations $\sim2\,\upmu$m, resulting in spin-exchange interaction strengths $J_\perp\sim40$\,Hz.
The experimental results are shown in Fig.~\ref{fig:tweezers-entanglement}(b) and Fig.~\ref{fig:tweezers-entanglement}(c) respectively.
Both groups prepared maximally-entangled Bell pairs and characterised the fidelity of entanglement: the maximum reported fidelities, after correcting state-preparation and readout (SPAM) errors, were $86.3(2.5)\%$ and $89(6)\%$ respectively.
These fidelities were primarily limited by single-particle decoherence and the finite temperature of the molecules.

Subsequent experiments extended these techniques to assembled RbCs~\cite{Ruttley2025} and NaCs~\cite{Picard2025} molecules, as shown in Fig.~\ref{fig:tweezers-entanglement}(d,e).  
Being assembled, these molecules were colder than the molecules used in the CaF experiments.
Further, in the case of RbCs, single-particle dephasing was effectively eliminated by using a magic-wavelength trap.
These two properties enabled higher SPAM-corrected entanglement fidelities of $97.6^{+1.4}_{-1.6}\%$\footnote{This notion means that the $1\sigma$ uncertainties on the measured value of $97.6\%$ are asymmetric: the upper uncertainty is $+1.4\%$ and the lower uncertainty is $-1.6\%$.} and $94(3)\%$, respectively.
In the case of RbCs, the entangled states had second-scale lifetimes, enabling quantum-enhanced metrology of external fields.
In the case of NaCs, the large dipole moment (see Table~\ref{Tab:1}) enabled sub-millisecond entanglement, which the team used to implement an iSWAP gate between hyperfine qubits. As before, the entanglement fidelities were primarily limited by motional dephasing caused by the finite molecular temperature.
Enhancing these entanglement fidelities will be critical to future work towards quantum computing with ultracold molecules.

\section{Conclusion}

The field of ultracold molecules has reached a very exciting stage. All the obstacles that have hampered progress have been overcome in the last few years. As we have shown, molecules can now be manipulated with the same level of control as atoms. Collisions can be controlled and gases of molecules can be cooled to quantum degeneracy. Quantum superpositions can be engineered with exceptionally long coherence times. Molecules can be controlled and detected at the single particle level, both in optical lattices and optical tweezers. Controlled long-range dipolar interactions can be introduced and used to engineer entanglement. With these developments, the field is now poised to move into an era of applications, harnessing the complexity of molecules to push beyond what is possible in other platforms.

Looking ahead, a key goal is to leverage the rich internal structure of molecules to its full advantage. For example, the internal states could be used to encode qudits rather than qubits~\cite{Sawant2020} or to go beyond the usual spin-$\frac{1}{2}$ paradigm~\cite{Hepworth2025}. Similarly, the rotational and hyperfine structure could be used to encode synthetic dimensions that extend to many synthetic lattice sites~\cite{Sundar2018}, leading to novel and unexplored physics when interactions between molecules are present~\cite{Sundar2019}. In lattices, the long lifetimes of molecular states should allow detailed examination of highly tunable $t-J$ models~\cite{Gorshkov2011}, chiral excitations \cite{syzranovSpinorbitalDynamicsSystem2014,syzranov_emergent_2016}, spin liquids \cite{yao_quantum_2018}, topological insulators \cite{yao_topological_2012,schusterFloquetEngineeringUltracold2021a} and much more. Controlling the lattice filling, particularly in quantum gas microscopes, is still challenging, but the advent of molecular BECs could be pivotal. Similarly, combining optical tweezers and optical lattices is a potentially interesting direction to explore with molecules offering a new level of control. There is also emerging work on combining molecules with Rydberg atoms~\cite{Guttridge2023,Zhu2025} to realise a hybrid quantum platform that offers opportunities for non-destructive readout of the molecular state and fast entangling operations~\cite{Zhang2022Rydberg,Wang2022}. Finally, we should not forget the importance of molecules for tests of fundamental physics~\cite{DeMille2024}. The staggering developments reported in this chapter also lay the foundation for a new generation of experiments using ultracold molecules, potentially harnessing dipolar interactions for quantum-enhanced measurements~\cite{Bilitewski2021,Wellnitz2024}.

\section{Acknowledgments}
The authors thank S. I. Mistakidis for a thorough reading of
this chapter. We acknowledge support from the UK Engineering and Physical Sciences Research Council (EPSRC) Grants EP/P01058X/1, EP/W00299X/1, and UKRI2226 funded through the Programme Grant Scheme, UK Research and Innovation (UKRI) Frontier Research Grant EP/X023354/1, the Royal Society, and Durham University. P.D.G. is supported by a Royal Society University Research Fellowship URF/R1/231274.

For the purpose of open access, the authors have applied a Creative Commons attribution (CC BY) licence to any Author Accepted Manuscript version arising.

This chapter will appear in the Springer book \textit{Short and Long Range Quantum Atomic Platforms – Theoretical and Experimental Developments} (provisional title) edited by P. G. Kevrekidis, C. L. Hung, and S. I. Mistakidis.

\bibliographystyle{spphys}
\bibliography{ref}

\end{document}